\documentclass[%
 aip,
 amsmath,amssymb,
 reprint,%
nofootinbib    
]{revtex4-1}

\usepackage{graphicx}
\usepackage{dcolumn}
\usepackage{bm}
\usepackage{caption}    
\usepackage{subcaption}    
\usepackage{array}  
\usepackage[hidelinks,pagebackref,colorlinks=true,allcolors=blue]{hyperref}	

\usepackage{xcolor}   
\usepackage{soul}   
\usepackage[export]{adjustbox}  

\usepackage[utf8]{inputenc}
\usepackage[T1]{fontenc}
\usepackage{mathptmx}
\usepackage{etoolbox}
\usepackage{environ}         

\DeclareUnicodeCharacter{2060}{!!!!I-AM-HERE!!!!!}   
\DeclareUnicodeCharacter{03BD}{!!!!I-AM-HERE!!!!!}   
\DeclareUnicodeCharacter{03C3}{!!!!I-AM-HERE!!!!!}   

\makeatletter
\def\@email#1#2{%
 \endgroup
 \patchcmd{\titleblock@produce}
  {\frontmatter@RRAPformat}
  {\frontmatter@RRAPformat{\produce@RRAP{*#1\href{mailto:#2}{#2}}}\frontmatter@RRAPformat}
  {}{}
}%
\makeatother

\newlength{\myl}
\let\origequation=\equation
\let\origendequation=\endequation

\RenewEnviron{equation}{
  \settowidth{\myl}{$\BODY$}                       
  \origequation
  \ifdimcomp{\the\linewidth}{>}{\the\myl}
  {\ensuremath{\BODY}}                             
  {\resizebox{\linewidth}{!}{\ensuremath{\BODY}}}  
  \origendequation
}

\begin{document}

\title{Effect of detachment on Magnum-PSI ELM-like pulses: II. Spectroscopic analysis and role of molecular assisted reactions}

\author{Fabio Federici}

\altaffiliation{E-mail address: fabio.federici@york.ac.uk}
\affiliation{ 
York Plasma Institute, Department of Physics, University of York, Heslington, York, YO10 5DD, United Kingdom
}%
\affiliation{ 
Oak Ridge National Laboratory, Oak Ridge, Tennessee 37831, USA
}%
\affiliation{ 
United Kingdom Atomic Energy Authority, Culham Centre for Fusion Energy, Culham Science Centre, Abingdon, Oxon, OX14 3DB, United Kingdom
}%

\author{Bruce Lipschultz}%
\affiliation{ 
York Plasma Institute, Department of Physics, University of York, Heslington, York, YO10 5DD, United Kingdom
}%

\author{Gijs R. A. Akkermans}%
\affiliation{ 
DIFFER-Dutch Institute for Fundamental Energy Research, De Zaale 20, 5612 AJ Eindhoven, The Netherlands
}%

\author{Kevin Verhaegh}%
\affiliation{ 
United Kingdom Atomic Energy Authority, Culham Centre for Fusion Energy, Culham Science Centre, Abingdon, Oxon, OX14 3DB, United Kingdom
}%
\affiliation{ 
York Plasma Institute, Department of Physics, University of York, Heslington, York, YO10 5DD, United Kingdom
}%

\author{Matthew L. Reinke}%
\affiliation{ 
Oak Ridge National Laboratory, Oak Ridge, Tennessee 37831, USA
}%
\affiliation{ 
Commonwealth Fusion Systems, Cambridge, MA 02139, USA
}%

\author{Ray Chandra}%
\affiliation{ 
DIFFER-Dutch Institute for Fundamental Energy Research, De Zaale 20, 5612 AJ Eindhoven, The Netherlands
}%
\affiliation{ 
Aalto University, Otakaari 24, 02150 Espoo, Finland
}%

\author{Chris Bowman}%
\affiliation{ 
United Kingdom Atomic Energy Authority, Culham Centre for Fusion Energy, Culham Science Centre, Abingdon, Oxon, OX14 3DB, United Kingdom
}%

\author{Ivo G. J. Classen}%
\affiliation{ 
DIFFER-Dutch Institute for Fundamental Energy Research, De Zaale 20, 5612 AJ Eindhoven, The Netherlands
}%

\author{Magnum-PSI Team}%
\affiliation{ 
DIFFER-Dutch Institute for Fundamental Energy Research, De Zaale 20, 5612 AJ Eindhoven, The Netherlands
}%


\begin{abstract}
The linear plasma machine Magnum-PSI can replicate similar conditions to those found in a tokamak at the end of the divertor leg. A dedicated capacitor bank, in parallel to the plasma source, can release a sudden burst of energy, leading to a rapid increase in plasma temperature and density, resulting in a transient heat flux increase of half of an order of magnitude, a so called ELM-like pulse. Throughout both the steady state and the pulse, the neutral pressure in the target chamber is then increased, causing the target to transition from an attached to a detached state. In the first paper related to this study\cite{Federici} direct measurements of the plasma properties are used to qualitatively determine the effect of detachment on the ELM-like pulse.
Here measurements from a purposely improved optical emission spectrometer were used in conjunction with other diagnostics to build a Bayesian algorithm capable of inferring the most likely properties of the plasma, poloidally and temporally resolved. 
This is used to show the importance of molecular assisted reactions. Molecular processes, and especially molecular activated dissociation, are found to be important in the exchange of potential energy with the plasma, while less so in radiating the energy from the ELM-like pulse. At low target chamber pressure, the plasma generated via ionisation during the part of the ELM-like pulse with the higher temperature is more than that produced by the plasma source, a unique case in linear machines. At high target chamber pressure molecular activated recombination contributes up to a third of the total recombination rate, contributing to the reduction of the target particle flux. Some metrics that estimate the energy lost by the plasma per interactions with neutrals, potentially relevant for the portion of the tokamak divertor leg below $\sim10eV$, are then tentatively obtained.

\end{abstract}

\maketitle


\section{Introduction}\label{introduction}

In tokamaks, the divertor region is crucial for managing the exhaust of high-temperature plasma. Current technology limits the heat flux to the divertor target, and mitigation methods are needed to reduce it to acceptable levels ($<10MW/m^2$). \cite{Pitts2019} One mitigation method is detachment, creating a low-temperature buffer in front of the target, which dissipates the plasma energy and particles via radiation and recombination. Detachment can be induced by increasing divertor density, seeding impurities, or reducing power flow. \cite{Leonard2018} Detachment has been demonstrated to significantly reduce target heat flux in experiments like Alcator C-Mod and AUG.\cite{Lipschultz2007,Kallenbach2015a}

In high performance core regimes, short bursts of heat and particles called edge localised modes, ELMs can increase heat flux to intolerable levels. \cite{Jachmich2011} When ELMs occur during detachment, the divertor plasma temporarily reattaches, helping to reduce target heat load by dissociating and ionizing neutrals. However, studying this dynamic process is challenging.

This study aims to investigate how ELM-like pulses interact with a detached target plasma, and to gain insights into the processes responsible for removing energy from the ELM before it reaches the target. During detachment the region in front of the target is cold so neutral and molecular hydrogen densities are high and thus the role of those species in removing power, momentum and particles is studied. Of particular interest is the relevance of molecular assisted processes (ionisation (MAI), recombination (MAR) and dissociation (MAD)) over atomic. This was studied before in tokamaks and linear machines but never during the ELM burn through. \cite{Akkermans2020,Verhaegh2021a} 
This work is also important because different phenomena are at play and have to be correctly understood to gain predictive capability for the ELM burn through in tokamaks. The filamentary nature of the tokamak ELM makes a 3D treatment necessary\cite{Smith2020,Smith2020a} while its fast transient nature makes kinetic effects relevant.\cite{Mijin2020} On top the interaction with cold neutrals, with solid surfaces and the cooling of the plasma to sub eV temperature require transport and a large number of interactions to be accounted.\cite{Zhou2022,Tskhakaya2009} The presence of molecular precursors like ${H_2}^+$ and $H^-$ and their interactions with plasma and neutrals further complicate the picture, so it is necessary to asses if they play a significant role in the burn through process.
This is the second paper deriving from this study. The first shows what can be inferred from direct measurements of the plasma properties (temperature, density, light emission) and interaction with the target (thermography).\cite{Federici}

To investigate consistently these phenomena, ELM-like pulses are reproduced in Magnum-PSI, a linear plasma device in the DIFFER laboratory, The Netherlands. Unlike a tokamak, Magnum PSI and other linear machines feature a simpler geometry, allowing for an easier interpretation, improved repeatability, and enhanced diagnostic access. Magnum-PSI is also capable of producing steady-state target perpendicular heat fluxes that are comparable to those expected at the ITER target. A capacitor bank (CB) is connected in parallel to the plasma source power supply and generates the ELM-like pulses, while increasing hydrogen neutral pressure in the target chamber induces detachment.
The Optical Emission Spectroscopy (OES) setup was improved to increase the time resolution in order to collect data on the ELM-like pulse behaviour. An analysis of the  power and particle balance in the plasma column inside the target chamber, separating molecular and atomic contributions to power and particle losses, was performed through a purpose built Bayesian routine that also makes use of the new OES data.

Linear machines are significantly different from the exhaust of tokamaks. In a tokamak divertor, especially in the high recycling regime preceding detachment, the main contributor to the ion target flux is not the flow from upstream but the ionisation of neutrals recycling at the target, while the upstream acts as a power source. 

The divertor leg, in terms of particle balance, acts more as a closed system being supplied with energy from the upstream boundary \cite{Krasheninnikov2017a}. In linear machines, instead, the upstream plasma flow usually dominates, and the ionisation source is minor. The plasma flows from the source to the target, where its temperature and density are usually too low to cause recycling.
Another difference is the connection length, of the order of tens of meters in tokamaks while it is less than one in Magnum-PSI. In tokamaks, the wave of hotter plasma due to the ELM can propagate along field lines, progressively burning through the barrier of cold neutrals generated by detachment. The temporal dynamics of the upstream conditions have comparable timescales to the transport of heat along the magnetic field lines\cite{Smith2020,Leonard1999}. In linear machines the flow from source to target is much faster than the evolution upstream, so that the discharge behaves more as a succession of steady states. In linear machines is difficult, therefore, to study the effect of the ELM burning through the gas buffer created by detachment, while they should be more representative of the cross field neutral transport.

The ELM frequency can be high in tokamaks, from fractions to tens of kHz. Considering recycling confines the neutrals at the target, it was postulated from simulations that the neutral pressure in the divertor could progressively increase as more ELMs happen,\cite{Smith2020} making it difficult to study a quasi steady state reference state. Due to the lower ELM frequency achievable in Magnum-PSI, and the low ion source, this effect is usually not observed and the effect of repeated ELMs can be studied with respect to the same steady state case. This enables delineating the impact of an ELM on a neutral gas buffer from changes of the neutral gas buffer due to previous ELMs interacting with the target. In tokamaks there is also a strong correlation between ELM frequency and energy released during the ELM,\cite{Jachmich2011} making the independent study of these two aspects difficult, while in linear machines these can be more easily controlled.

The presence of impurities likely plays a role in the interactions of ELMs with detachment, both due to increased radiative and ionisation power losses, and because of the increased sputtering when reaching the target. This is not guaranteed to be reproduced in linear machines unless specific setups are arranged.
Linear machines, despite these differences, remain a valuable tools in understanding the effects of neutral pressure on tokamak ELMs and what processes are involved.

In \cite{Federici} it was demonstrated that increasing the neutral pressure within the target chamber results in removing energy from the ELM-like pulses. In some cases, the ELM-like pulses have no effect on the target temperature (Section V). The target heating from the ELM-like pulse is comparable to measurements from current tokamaks, although considerably lower than what is anticipated in larger scale devices like ITER. Furthermore, this can be mitigated by raising the neutral pressure. Observations from the visible light fast camera (Section III) were also used to divide the interaction of the ELM-like pulses in phases depending on the detachment in steady state and during the ELM. In Stage 1 the plasma is attached to the target both in steady state and during the ELM. In Stage 2, the target is cold in between ELM-like pulses but a significant target heating can be provided transiently. In Stage 3 there is no significant heat transferred to the target the target either before or during the ELM-like pulse.
The visible light emission is not homogeneous in the whole target chamber, with a peak close to the target. The emission in the bulk of the target chamber quite uniform and decreases at the target as the density increases. This observation will be used later to support the approximation that volumetric power losses can be considered constant from skimmer to target.

It will here be shown that for increasing neutral pressure, the energy and particles of the ELM like pulse will increasingly be removed in the Magnum-PSI target chamber volume and that an important role is played by plasma-molecule interactions. The radiated power losses are a significant power loss channel, but elastic collisions with neutrals and exchanges of potential energy dominate in reducing the plasma temperature to levels where recombination becomes important.



\section{Experimental conditions}\label{Experimental conditions}

\begin{figure*}
	\centering
	\includegraphics[width=\linewidth,trim={30 0 0 0},clip]{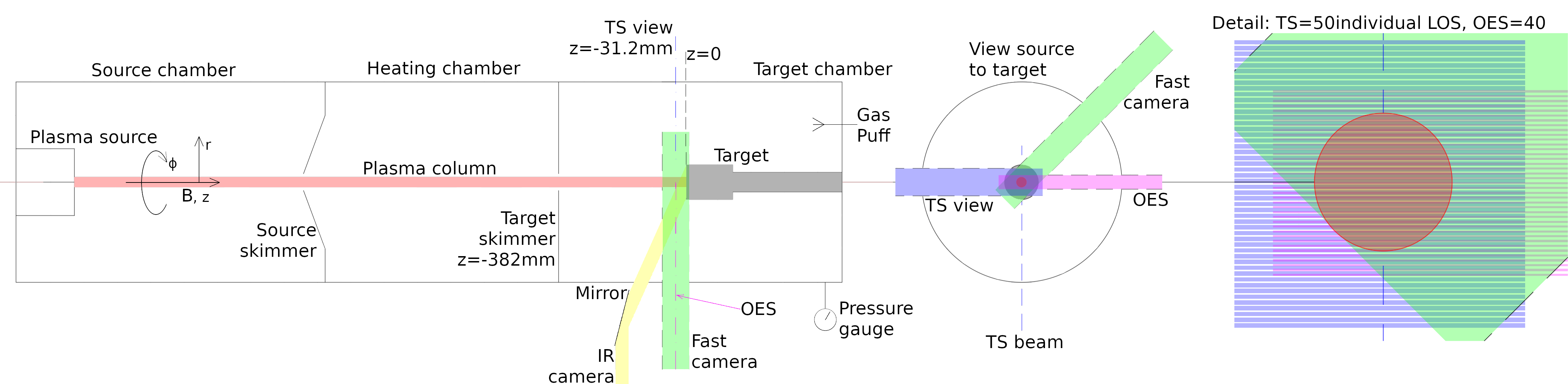}
	\caption{Diagram of Magnum-PSI linear plasma device. The target's surface is at $z=0$, the Lines of Sight (LOS) Thomson scattering (TS) and Optical Emission Spectroscopy (OES) are both at $z=-31.2$mm. The TS view consists of 50 evenly spaced LOS that are perpendicular to the plasma column length ($z$ axis). The OES view comprises 40 similarly oriented LOS. The fast camera, that measures the line integrated light emission from the plasma, has a radial view of the plasma column through a viewport on the side of the vacuum vessel of diameter $\sim$75mm. The IR camera view is also radial, but is converted to aim on the target surface via a mirror located outside the target chamber. The mirror's angle is adjusted when the target is moved in $z$, to keep the target into view.}
	\label{fig:layout}
\end{figure*}

A diagram of the Magnum-PSI device is shown in \autoref{fig:layout}.

\autoref{tab:table1} shows the portion of experimental conditions used on Magnum-PSI for this study that will be analysed in this paper, termed strong pulses. An additional set of pulses characterized by a weaker magnetic field and lower energy stored in the capacitor banks, referred to as weak pulses, also exists but is analysed in this paper. For weak pulses the plasma temperature becomes so low that the TS measurement only works at the peak of the ELM-like pulse and is not representative of the entire plasma column. Consequently, weak pulses are unsuitable for analysis using the Bayesian model.
For a more detailed description of the discharge procedure and the sampling strategy, please refer to Appendix A of \cite{Federici}.


\begin{table*}
\begin{tabular}{ | >{\centering}m{01em} | >{\centering}m{1.5cm}| >{\centering}m{1.5cm} | >{\centering}m{1.2cm} | >{\centering}m{2.2cm} | >{\centering}m{2.2cm} | >{\centering}m{3.6cm} | >{\centering}m{2cm} | } 
  \hline
  ID & capacitor voltage [V] & capacitor energy [J] & Magnetic field [T] & Target chamber $H_2$ feeding [slm] & Target chamber pump speed [\%] & Steady state neutral pressure in target chamber [Pa] & Stage\cite{Federici} \tabularnewline 
  \hline
  5 & 800 & 48.0 & 1.3 & 0 & 82 & 0.296 & 1\tabularnewline
  \hline
  6 & 800 & 48.0 & 1.3 & 0 & 25 & 0.516 & 1\tabularnewline
  \hline
  7 & 800 & 48.0 & 1.3 & 5 & 25 & 4.370 & 1/2\tabularnewline
  \hline
  8 & 800 & 48.0 & 1.3 & 10 & 25 & 8.170 & 2\tabularnewline
  \hline
  9 & 800 & 48.0 & 1.3 & 15 & 25 & 11.847 & 2\tabularnewline
  \hline
  10 & 800 & 48.0 & 1.3 & 20 & 25 & 15.040 & 2\tabularnewline
  \hline
\end{tabular}
  \caption{Table of Magnum-PSI experimental conditions for the experiments analysed in this paper, referred to as strong pulses. The most important parameters common for all cases are: steady state source current 140A (to generate the background plasma), 31.2mm OES/TS to target distance,TZM (a molybdenum alloy) target. The stages go from 1 to 3, corresponding to: fully attached (1), partially detached (2) and fully detached (3).\cite{Federici}}
  \label{tab:table1}
\end{table*}

\section{Diagnostics}\label{Diagnostics}

In this section we will be describe  the diagnostics utilised in this study. While some have the time resolution required to observe the evolution of individual ELM-like pulses, others can observe only a limited portion. Therefore, the sampling strategy adopted allows to reconstruct the full ELM-like pulse evolution.

The diagnostics used in this paper are:
\begin{itemize}
    \item Thomson scattering (TS): electron temperature and density
    \item Jarell-Ash, Czerny-Turner spectrometer OES: Hydrogen atomic line emission (Balmer series $p=4-8 \rightarrow 2$)
    \item Power source (ADC): temporal variation of the power delivered to the plasma
\end{itemize}

\subsection{Optical emission spectrometer (OES)}\label{Optical emission spectrometer}

The main component is a Jarell-Ash spectrometer connected to a fibre optic bundle with 40 individual cores that view the plasma radially with a spatial resolution of 1.06mm, individual line of sight (LOS) width of $\sim$1mm. See the OES LOS in the target chamber in \autoref{fig:layout}, further detail from Barrois. \cite{Science2017}

Before this work was conducted the camera connected to the spectrometer was a Princeton Instruments PIXIS 2048B, with a shutter speed of the order of seconds. In order to obtain information on the behaviour during the ELM-like pulse this camera was replaced in 2019 by the author and Gijs Akkermans with a Photometrics Prime95B 25mm RM16C with a minimum integration time of $20\mu s$. 

The camera has a CMOS sensor with rolling shutter, meaning the exposure of one row happens after the previous row is completed, forcing the accumulation of data on multiple ELM-like pulses to reconstruct the full spatial and temporal brightness profile of the pulse. Details on how to sample all stages of the ELM-like pulse so that one obtains a coherent picture are shown in Appendix A of \cite{Federici} while the steps from raw images to line emissivity are detailed in \ref{OES data interpretation}. The sensitivity calibration was done using a Labsphere as explained extensively by Barrois. \cite{Science2017}

During the experiments hydrogen Balmer lines brightness ($p=4-\infty \rightarrow 2$) were recorded; only lines $p=4-8 \rightarrow 2$ were considered to have a sufficient signal to noise ratio and were used in this study.


\subsection{Thomson scattering (TS)}\label{Thomson scattering1}
The Thomson scattering (TS) diagnostic allows the measurement of both $T_e$ and $n_e$. A laser beam is fired through the plasma and the scattered light is  collected from the measurement volume determined by the intersection of the laser beam with each TS viewing LOS. 50 LOS are simultaneously measured in a radial plane (the same as OES, as shown in \autoref{fig:layout}), to find the profile of the plasma properties. In Magnum-PSI, TS is used for both steady state plasmas and, albeit with reduced performance, time dependent measurements. For the experiments part of this study, TS was used in time dependent mode, with $50\mu s$ time resolution and integration time. This returns an uncertainty $<3\%$ for $n_e$, and $<10\%$ for $T_e$ for $n_e>2.8 \cdot 10^{20} \#/m^3$. As time between consecutive measurements has to match the laser frequency of 10Hz, we were forced to accumulate data over multiple pulses to reconstruct the entire ELM-like pulse time evolution. \cite{VanDerMeiden2012}


\subsection{Power supply}\label{Power supply}
The steady state power supply regulates the DC plasma source voltage while a capacitor bank composed of 28 individual sections is connected in parallel to cause a strong  increase of the plasma source current. The energy stored in each capacitor (shown in \autoref{tab:table1}) is given by $\frac{1}{2}CV^2$. 
During ELM-like pulses, 92\% of the electrical energy dissipated in the plasma source is converted into plasma energy.\cite{Morgan2014} Additionally, Some energy is dissipated before the plasma reaches the target skimmer within the source and heating chambers. Here the pressure is maintained as low as possible via differential pumping, thereby reducing the interaction of the plasma column with the cold gas in these chambers and minimizing the associated energy losses. 
From the increase in cooling water temperature in steady state, it is estimated that the source and target skimmers absorb $\sim$10\% and $<$2\% of the electrical input energy, respectively.

To gain more insight than what can be gained by individual diagnostics, presented in \cite{Federici}, the OES data will be combined with that from TS and the input power source measurement within a Bayesian analysis framework. The purpose is to identify which processes are more relevant in the increase of energy removed in the volume for increasing levels of steady state detachment, driven by target chamber neutral pressure, and possibly their evolution in time.

\section{Bayesian analysis}\label{Role of molecular assisted reactions}
In this section we describe our analysis technique, that simultaneously takes into account information from different diagnostics on Magnum-PSI data to gain more insight into the dominant atomic and molecular processes during an ELM-like pulse. A significant aspect of this analysis is the inclusion of the molecular species $H_2$, ${H_2}^+$ and $H^-$. These species interact with the plasma, leading to the formation of excited hydrogen atoms, with related power and particle sources/sinks, which have a significant influence on the behaviour of Magnum-PSI target chamber plasmas, driving them into detachment. Since the interaction of these species with the plasma causes hydrogen atomic line emission they will be referred to as molecular precursors.

The ordinary method for inferring the plasma properties from Balmer emission lines is to consider that the emission from higher-p H excited states is mainly generated via electron-ion recombination (EIR), while the emission from lower excited states is generated via electron impact excitation (EIE), and molecular contributions are generally neglected for simplicity. This normally allows to determine if the plasma is ionising, recombining or in between.\cite{Potzel2012,Verhaegh2019}. Under conditions like those in this work, on the other hand, molecular contributions can be significant.\cite{Akkermans2020} Plasma molecular interactions lead to excited hydrogen atoms generally excited to low $p$ levels, and their distribution across p is similar to that of EIE, thus complicating the Balmer line analysis (see \autoref{Limitations due to the measured lines}).

A Bayesian analysis, that employs a probabilistic approach, is used to determine the plasma/molecule/atom characteristics ($T_e$, $n_e$, $n_{H}/n_e$, $n_{H_2}/n_e$, $n_{{H_2}_+}/n_{H_2}$, $n_{{H}^-}/n_{H_2}$, one of these quantities is referred to as $\Theta_i$, while a set of 6 as $\Theta$) that best match both the line emission (from OES) and the electron temperature and density (from TS). This approach combines information from different diagnostics, incorporates priors from simulations as well as reaction rates to find the best match to experimental data, providing insights into the dominant processes and their effects on the energy and particle balance.
The main advantage of this approach is that it can not only account for all uncertainties and non linear rates, but also allows calculating the probability density function (PDF) of the quantities of interest. This implies that multiple non unique solutions, if present, can be found.

Probabilistic approaches have been used before to infer the full state of the divertor plasma in tokamaks, but never during the ELM burn through.\cite{Verhaegh2017,Verhaegh2020,Bowman2020} First, the parameter space of $\Theta$ is defined, forming a regular grid. The range and samples of the $\Theta_i$ parameter are defined based on Thomson scattering and other priors. Given a certain combination of parameters $\Theta$ the outputs of all the quantities of interest can then be calculated, like line emission, power and particle sinks and sources, etc. Those predicted observables can then be compared with the priors to find the probability that the calculated values match the expected ones. This comparison and the calculation of the observables depend on the specifics of the problem and the simplifications used. For the combination of parameters $\Theta$ the probability from the priors is multiplied with the probability from the comparison of the calculated and measured line emission, returning the likelihood of the particular $\Theta$. Starting with this grid, one can either add individual $\Theta$, searching for the maximum of the likelihood, or refine the grid further around the best $\Theta$. A grid refinement is done here to retain a grid structure and simplify the numerical problem. This is done only twice to limit the computational cost and memory requirements. After the quantities of interest like, for example, terms of the power balance, can then be calculated for each $\Theta$ and their probability density function (PDF) found.


\autoref{Balance over the plasma column} gives the full list of assumptions used to simplify the physics of the plasma of interest (from the target skimmer to the target, see \autoref{fig:layout}). In short, the plasma is considered homogeneous along its length ($z$) from the target chamber skimmer to the target within each time step, and being poloidally ($\phi$ in \autoref{fig:layout}) symmetric for a given radius ($r$). As mentioned before this is supported by the relatively uniform visible light emission far from the target found in Stage 1 and 2.\cite{Federici} This means that the ELM-like pulse is not localised at a particular spatial location but happens along the whole plasma column, matching well the fast nature of the transport along field lines ($\sim$ sonic flow, $<20 \mu s$ to flow through the target chamber) compared to the changes in plasma conditions over time (changes well resolved with $50 \mu s$ time steps). The conditions at the target chamber skimmer will be considered as the input to the plasma column. The presence of impurities is monitored by a 6-channels Avantes AvaSpec-2048-USM2-RM survey optical emission spectrometer covering the range 299-950nm.\cite{Science2017} The survey spectrometer's channels LOSs are in the target chamber and impurities like metals from the power source or oxygen from cooling water are only detectable in significant amounts when the plasma source fails. This shows that impurities from the source are ordinarily efficiently removed in Magnum-PSI by differential pumping. Carbon can play a significant role when graphite targets are used\cite{VanEck2013}, but this is not a concern in these experiments in which a TZM target was used.

The integral over the ELM-like pulse of quantities of interest, like terms of the power balance, are obtained by summing the contribution from each time step and radial location. In the following section we will explain the method used to determine the probability associated with an output at a given time and radial location.

\subsection{Analysis steps}\label{Analysis steps}

We will describe here in more detail the steps to first initialize the paramenter space based on TS and other priors (\autoref{Parameter space initialisation}). Then we will show how to compare the modelled quantities and TS with the priors to  obtain the probability of each $\Theta$ (\autoref{Prior probability distribution}). Lastly the comparison with the measured Balmer lines emission, grid refinement process and determination of the quantity of interest PDFs will be illustrated (\autoref{Bayesian analysis}). The numbering of each step corresponds with the numbers in the boxes in \autoref{fig:bayes1a}, \ref{fig:bayes1b}, \ref{fig:bayes1c}.

\subsubsection{Parameter space initialisation}\label{Parameter space initialisation}

\begin{figure}
	\centering
	\includegraphics[scale=0.29,trim={0 0 0 0},clip]{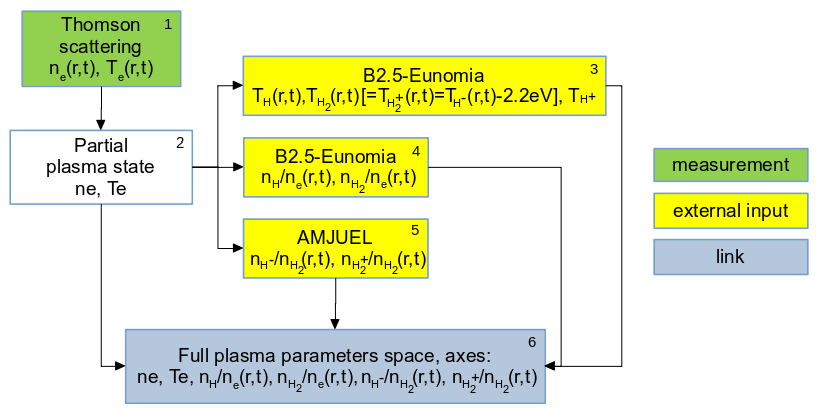}
	\caption{Sketch illustrating how the initial guess of each parameter $\Theta_i$ is done. With \emph{link} are marked items that link different parts of the global bayesian routine together. \emph{measurement} indicate quantities measured experimentally and \emph{external input} represents inputs from simulations or reaction rates libraries.}
	\label{fig:bayes1a}
\end{figure}


Our Bayesian analysis first finds the range of parameters to utilize. This is shown in \autoref{fig:bayes1a}. Every point in time and radius $(t,r)$ during the ELM-like pulse is considered independently. Referring to \autoref{fig:bayes1a} the meaning of the steps are as follows:
\begin{enumerate}
    \item[1,2] \emph{Thomson scattering $n_e$(r,t), $T_e$(r,t)}: \\The measured $T_e$ and $n_e$ and their uncertainties are used to specify the range and grid elements of $T_e$ and $n_e$. From every $T_e,n_e$ combination the samples for the other $\Theta_i$ are calculated.
    \item[4] \emph{B2.5-Eunomia $n_H/n_e$(r,t), $n_{H_2}/n_e$(r,t)}: \\ We utilize B2.5-Eunomia simulations, a multi-fluid plasma code coupled with a non-linear Monte Carlo transport code for neutrals\cite{Wieggers2013,Chandra2021,Chandra2022}, to determine scalings for $n_{H_2}$ and $n_H$ with $T_e$. We utilize these scalings to obtain ranges for $n_{H_2}/n_e$ and $n_H/n_e$ (see \autoref{Priors from B2.5 Eunomia} and \ref{Priors range optimization}).
    \item[3] \emph{B2.5-Eunomia $T_H$(r,t), $T_{H_2}$(r,t)[$=T_{{H_2}^+}(r,t)=T_{H^-}(r,t)-2.2eV$], $T_{H^+}$}: \\Similarly, values of $T_H$, $T_{H_2}$ and $T_{H^+}$ are extracted from B2.5-Eunomia simulation scalings with $T_e$. Given the molecular precursor ${H_2}^+$ is predominantly generated from $H_2$, it is assumed that $T_{H_2}=T_{{H_2}^+}$. In the creation of $H^-$ from $H_2$ the ion can get some of the Frank-Condon energy of the $H_2$ bond (2.2eV per nucleon). For this reason 2.2eV are added to $T_{H_2}$ to estimate $T_{H^-}$\cite{Verhaegh2020} (see \autoref{Priors from B2.5 Eunomia}).
    \item[5] \emph{AMJUEL $n_{H^-}/n_{H_2}$(r,t), $n_{{H_2}^+}/n_{H_2}$(r,t)}: \\${H_2}^+/H_2$ and $H^-/H_2$ density ratios are calculated from the AMJUEL collection of cross sections, reaction rates and ratios \cite{Reiter2017,Reiter2005,Kotov2007} to determine the $n_{H^-}/n_{H_2}$ and $n_{{H_2}^+}/n_{H_2}$ ranges used as priors (see \autoref{Priors from AMJUEL} and \ref{Priors range optimization}).
\end{enumerate}

\subsubsection{Prior probability distribution}\label{Prior probability distribution}

\begin{figure*}
	\centering
	\includegraphics[scale=0.29,trim={0 0 0 0},clip]{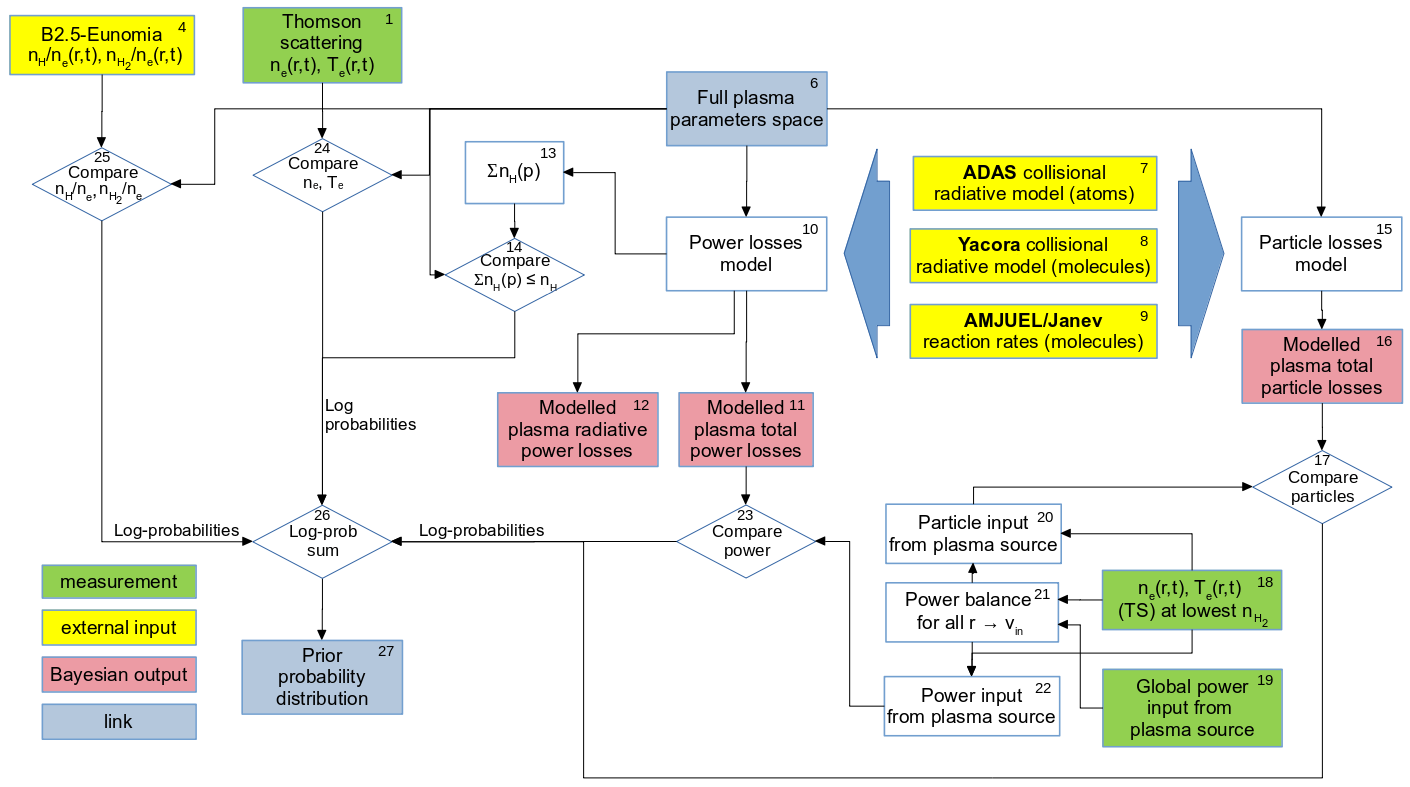}
	\caption{Sketch illustrating how the prior probability distribution is determined given the parameter grid. \emph{Bayesian output} indicates groups of quantities that are collected for all the parameter space and with the likelihood will return their PDF.}
	\label{fig:bayes1b}
\end{figure*}

Once the initial parameter space is determined, the prior probability distribution is calculated. The process for this is shown in \autoref{fig:bayes1b} and the meaning of the steps is as follows:

\begin{enumerate}
    \item[7] \emph{ADAS colisional radiative model (atoms)}: \\The Atomic Data and Analysis Structure (ADAS) collisional radiative model\cite{Summers2004,OMullane2013} is used to calculate the line emission (see \autoref{Emissivity}), the power losses (see \autoref{Power balance}) and the particle balance sources/sinks (see \autoref{Particle balance}) due to atomic processes. The reactions considered and relative coefficients are indicated in \autoref{tab:adas}. This type of analysis is fairly standardised, see for example \cite{Verhaegh2018}.
    \item[8] \emph{Yacora collisional radiative model (molecules)}: \\The Yacora collisional radiative model\cite{Wunderlich2016,Wunderlich2020} is used to calculate the line emission due to molecular precursors via the population coefficients (see \autoref{Emissivity}) and the radiative power losses (see \autoref{Power balance}). The molecular precursors that will be utilized in this analysis are $H_2$, ${H_2}^+$ and $H^-$ based on the work by Verhaegh \cite{Verhaegh2021} (see the reactions in \autoref{tab:yacora}). ${H_3}^+$ could also be considered, but its relevance is expected to be much lower than that of ${H_2}^+$ and $H^-$ so it is dropped to limit the number of variables.\cite{Verhaegh2021}.
    \item[9] \emph{AMJUEL/Janev reaction rates (molecules)}: \\The AMJUEL database and a collection of reaction rates from Janev \cite{Janev2003} is used to calculate the rate of reactions involving molecular precursors (see \autoref{tab:amjuel}, \ref{tab:janev}). The reactions rates are use then to calculate sources and sinks in the power and particle balance in the plasma column (see \autoref{Power balance}, \ref{Particle balance})
    \item[19] \emph{Global power input from plasma source}: \\The power input from the plasma source is calculated from the measurement of voltage and current on the plasma source. From a study  by Morgan\cite{Morgan2014}, 92\% of the electrical energy is transferred to the plasma during an ELM-like pulse. This value will be considered valid for each time step.
    \item[18] \emph{$n_e(r,t), T_e(r,t)$ (TS) at lowest target chamber pressure}: \\It is assumed that for the lowest pressure case the interactions between the plasma column and the background gas in the target chamber are negligible, therefore they represent the input conditions for all cases that differ only by the target chamber neutral pressure. They are referred to as $n_{e,in}, T_{e,in}$ and used to determine the input power and particle profiles (see \autoref{Balance over the plasma column}).
    \item[21] \emph{Power balance for all $r \rightarrow v_{in}$}: \\Assuming that the plasma flows from skimmer to target with the same Mach number ($M_{in}(t)$) across different radii it is possible to relate the total input power ($P_{source}(t)$) with the input conditions ($n_{e,in}(r,t), T_{e,in}(r,t)$). It is then possible to determine $M_{in}(t)$ and then $v_{in}(r,t)$. See \autoref{Balance over the plasma column} for details.
    \item[22] \emph{Power input from plasma source}: \\The maximum power input is calculated as the energy flow due to a plasma in the input conditions ($n_{e,in}(r,t), T_{e,in}(r,t)$) at flow velocity $v_{in}(r,t)$ plus the energy associated to the depletion of the plasma at the specific neutral pressure of the experiment ($n_e(r,t), T_e(r,t)$) in one time step. See \autoref{Power balance} for details.
    \item[20] \emph{Particle input from plasma source}: \\Similarly to the previous step the maximum particle input is obtained as the input particle flow using $n_{e,in}(r,t)$ and $v_{in}(r,t)$ plus the particle flow associated with the dissipation of the plasma at the specific neutral pressure of the experiment ($n_e(r,t)$) in one time step. See \autoref{Particle balance} for details.
    \item[10] \emph{Power losses model}: \\ The power losses in the annular section of the plasma column of interest are modelled as per \autoref{Power balance}. The effect of impurities is not considered, as it is assumed that most of the impurities introduced by the plasma source anode and cathode are removed by differential pumping in the source and heating chamber.
    \item[12] \emph{Modelled plasma radiative power losses}: \\The total radiated power from the plasma column is calculated. For atomic processes ADAS coefficients are used. For molecular processes Yacora coefficients are used, summing the line emission for all possible transitions (the highest excited state considered is p=13). It is not considered the radiation caused directly by $H_2, {H_2}^+, H^-$ excited states such as the Werner and Lyman bands, as it should have a negligible contribution.\cite{McLean2019,Groth2019}
    \item[11] \emph{Modelled plasma total power losses}: \\The total power losses are obtained by adding: total radiated power losses, net difference of potential energy from reactants to products and energy loss attributed to the heat carried away from the neutral produced by recombination. This terms and their components are part of the outputs of the Bayesian analysis.
    \item[23] \emph{Compare power}: \\The modelled net power losses are compared to the power input previously inferred. The net power losses have to be between 0 and the power input. The probability that this is the case is calculated as shown in \autoref{Power balance} and is one component of the prior density distribution.
    \item[15] \emph{Particle losses model}: \\The particle sink-sources terms are modelled as per \autoref{Particle balance}. We neglect cross field transport when analysing the particle balance. Given the difficulty of accounting for cross field transport only the particle balance of the charged particles ($e^-,H^+,{H_2}^+,H^-$) that are bound by the magnetic field is operated.
    \item[16] \emph{Modelled plasma total particle losses}: \\The net particle sink is calculated for $e^-,H^+,{H_2}^+,H^-$). These terms and their components are part of the outputs of the Bayesian analysis.
    \item[17] \emph{Compare particle}: \\The modelled net particle losses are compared to the particle input previously inferred. Bounds for physical values of the net particle losses are obtained from the particle input previously inferred as detailed in \autoref{Particle balance}. The probability that the particle sinks are within the bounds is calculated and their log-probability summed.
    \item[13] \emph{$\Sigma n_{H(p)}$}: \\Within the power losses model the total density of excited states is calculated. ADAS PEC coefficients are used for atomic processes while Yacora coefficients are used for molecular, summing the density for all excited states.
    \item[14] \emph{Compare $\Sigma n_{H(p)} \leq n_{H}$}: \\It is checked that the total density of excited states is lower than the density of atomic Hydrogen. At the temperature of interest in this work the density of excited states is always a small fraction compared to the ground state, so a probability of 1 is assigned if the condition is respected and 0 otherwise.   
    \item[24] \emph{Compare $n_e$, $T_e$}: \\The measured $T_e$ and $n_e$ and their uncertainties are used to assign the relative probability on the $n_e$ and $T_e$ axis of the parameter space, uniform on the others.
    \item[25] \emph{Compare $n_H/n_e$, $n_{H_2}/n_e$}: \\The previously mentioned scalings from B2.5-Eunomia simulations are used to assign weakly varying probabilities on the $n_H/n_e$, $n_{H_2}/n_e$ axis of the parameter space, being uniform on the others, as indicated in \autoref{Priors from B2.5 Eunomia}.
\end{enumerate}

\subsubsection{Bayesian analysis}\label{Bayesian analysis}

\begin{figure}
	\centering
	\includegraphics[scale=0.29,trim={0 0 0 0},clip]{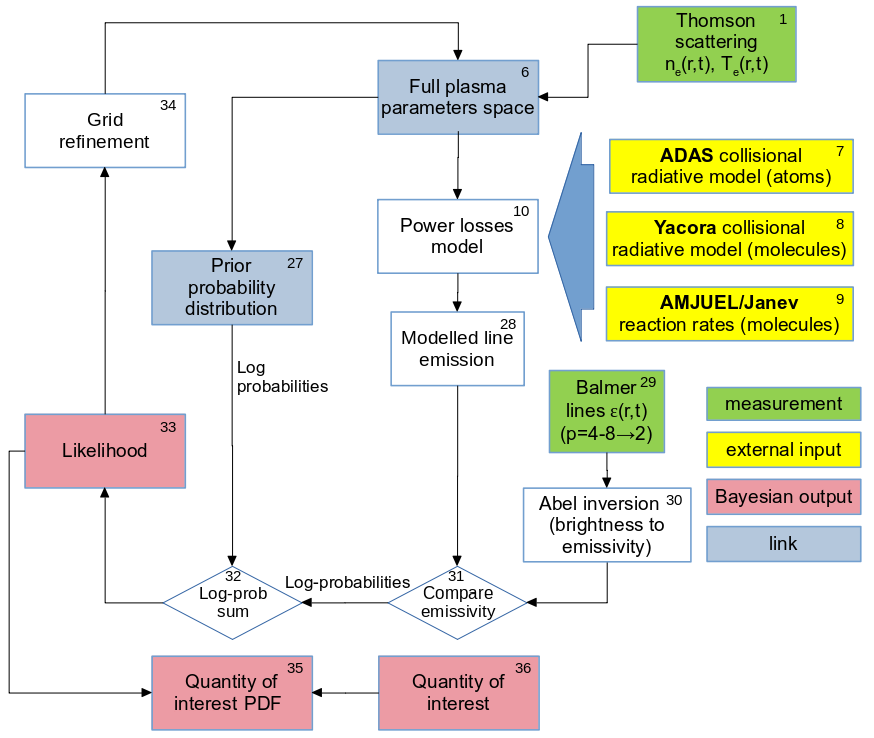}
	\caption{Sketch illustrating how the likelihood distribution is determined comparing the line emission with the expected values and adding the prior log-probability, to refine the parameter grid and calculate the outputs PFD.}
	\label{fig:bayes1c}
\end{figure}

The expected value of the Hydrogen line emissivity is then compared with the measured one, and the resulting probability combined with the prior to return likelihood, also referred to as the sum of the Log-probabilities. The grid is refined twice around the regions of the grid with high likelihood to improve the resolution of the probability distribution. While keeping a grid structure requires a large amount of memory the numerical procedure is relatively simple.
This procedure is shown in \autoref{fig:bayes1c} and the meaning of the blocks here introduced is as follows:

\begin{enumerate}
    \item[29] \emph{Balmer lines $\epsilon(r,t)(p=4-8\rightarrow 2)$}: \\The Balmer line emission for transitions $(p=4-8\rightarrow 2)$ is measured with OES (see \autoref{Optical emission spectrometer}). The camera has a rolling shutter so it is necessary to decouple the line of sight from the time information, as detailed in Appendix A of \cite{Federici} and \ref{OES data interpretation}.
    \item[30] \emph{Abel inversion (brightness to emissivity)}: \\Once the brightness per time step and line of sight is obtained an Abel inversion is operated (see \autoref{OES data interpretation}). The process converts the line integrated information to the local emissivity assuming poloidal symmetry and the plasma optically thin.
    \item[28] \emph{Modelled line emission}: \\As part of the power losses calculation the line emission from atomic and molecular processes the total line emission for the lines of interest ($p=4-8 \rightarrow 2$) is found (see \autoref{Emissivity}).
    \item[31] \emph{Compare emissivity}: \\The modelled line emissivity is compared to the measured one to return the probability that each combination of priors generated the measured emission (see \autoref{Emissivity}).
    \item[32,33] \emph{Likelihood}: \\The likelihood for every point of the parameter space is found adding the log probability from the prior and the comparison of the calculated emissivity with the measured one.
    \item[34] \emph{Grid refinement}: \\To increase the resolution in the region of the parameter space with high likelihood, the grid is refined by adding intermediate steps to the grid elements previously defined for every axis around the marginalised likelihood peak. The grid structure is maintained but it becomes non uniform. The prior probability distribution and comparison with measured emissivity is repeated to recalculate a new likelihood distribution. This loop is repeated twice to limit the computer memory requirements.
    \item[35,36] \emph{Quantity of interest PDF}: \\The quantity of interest for the plasma column power or particle balance have already been calculated for all the parameter space within \autoref{Prior probability distribution}. In order to reduce the size of the outputs the full range is reduced to a smaller number of intervals and the likelihood is summed within each interval.
\end{enumerate}

Once the PDFs for a quantity of interest in all radial and spatial locations are determined they are then convolved over radial steps, to obtain the total over the plasma column, and then in time over the ELM-like period to obtain the global quantities (see details in \autoref{Plasma column power balance}).

When a reaction occurs, energy is transferred between different species (e.g. plasma to molecules or atoms). If charged particles are created they are bound by the magnetic field and are confined to the same radial section of the plasma column. When neutral hydrogen atoms and molecules are generated they can escape the plasma column, carrying with them their kinetic energy. They can also interact with plasma at a different radii on the way out and  transfer the energy back in the plasma.

To model the transfer of energy to the neutrals and back to the plasma would require an effort beyond the scope of this work, so for now this component is not considered in the power and particle balance and only hyrdrogenic radiation is considered as net power leaving the plasma column. For more details see \autoref{Plasma column power balance}. 

It is important to address whether ignoring the power removed by the plasma column (and possibly transferred back) due to charge exchange (CX) and elastic collisions between $H_2$ and $H^+$ ($H_2$ elastic) is important and compatible with the Bayesian algorithm results. To estimate CX and $H_2$ elastic power losses we post-process our results as shown in \autoref{Power balance}. The same methodology was employed to reprocess B2.5-Eunomia results, and it returned a CX contribution between a 1\% and 150\% and $H_2$ elastic within 18\% and 34\% of the correct (self consistent) values. In one instance, the simplified method produced a factor of 100 larger CX losses compared to B2.5-Eunomia result. It is observed that in this case, with both high temperature and density, a large fraction of the CX losses are recovered by fast neutrals exchanging back to the plasma their energy before escaping the plasma column. This behaviour can be captured by the Monte Carlo model but not by this simplified method. With these very large uncertainties it is possible to say that CX is likely more important than $H_2$ elastic for ELM-like pulses, contrary to what shown in \cite{Chandra2022}, but the relevance for the plasma column energy balance is uncertain and would require a more detailed analysis, out of the scope of this work.

\subsection{Analysis limitations due to restrictions on measured Balmer lines}\label{Limitations due to the measured lines}
While it is in general desirable to measure as many transition lines as possible, it became apparent during the setup of the experiment that the H$\alpha$ line ($p=3\rightarrow 2$) could not be measured simultaneously with lines $p\geq 4\rightarrow 2$ due to the grating used. We describe herein how the lack of the H$\alpha$ line prevents the separation of the line emission contribution from ${H_2}^+$ and ${H^-}$ precursors.

\begin{figure}
	\centering
	\includegraphics[width=\linewidth,trim={0 0 0 36},clip]{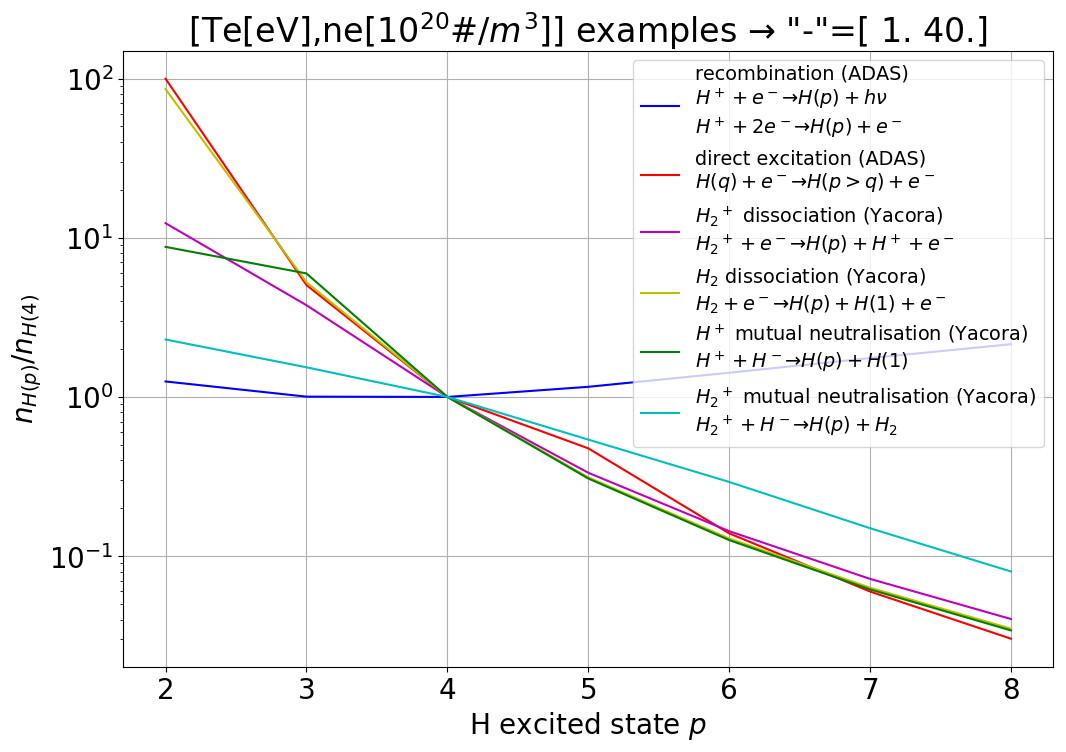}
	\caption{Density ratio of hydrogen excited states to p=4 for different processes for typical $T_e=1eV,n_e=4 \cdot 10^{21}\#/m^2$ plasma. The similarity of the profile shape for ${H_2}^+$ dissociation and $H^+$ mutual neutralisation is maintained for all the $T_e,n_e$ range of interest.}
	\label{fig:lines1}
\end{figure}

As a reference, \autoref{fig:lines1} shows the density ratio of excited states produced by different processes with respect to p=4 for typical $T_e,n_e$ values. 
Restricting Balmer lines to $n \geq 4 \rightarrow 2$ recombination, ${H_2}^+$ mutual neutralisation and EIE have a distinctively different profile while ${H_2}^+$ dissociation, $H_2$ dissociation and $H^+$ mutual neutralisation are fairly similar. $H_2$ dissociation is usually not problematic because the $H_2$ density required to match the measured emissivity is excessively high. Conversely, ${H_2}^+$ dissociation and $H^+$ mutual neutralisation can be caused by relatively low molecular precursors densities, making it difficult to distinguish which caused the emission. This could have been alleviated by including $p=3 \rightarrow 2$ in the analysis\cite{Verhaegh2021a} but this was not possible with the available gratings. \autoref{fig:lines1} illustrates that with the present setup, using only the line ratios available is insufficient to distinguish which molecular precursor, ${H_2}^+$ or $H^-$, is dominant. This could be alleviated by combining matching line intensities with other conditions as described in \autoref{Analysis steps}, as it can help rule out regimes that can well match the emission profile but would be unphysical.

The results of this study and the relevance of molecular interactions during the ELM-like pulse will now be shown.

\subsection{Bayesian analysis results}\label{Bayesian analysis results}
In this section the results from the Bayesian analysis will be presented. Only results regarding the strong pulse cases are shown, since the temperature and density falls below the detection capabilities of the TS system for a significant fraction of the duration for weak pulses.

\subsubsection{Power balance}\label{Power balance bayesian}
\begin{figure}
	\centering
     \begin{subfigure}{0.8\linewidth}
    	\includegraphics[width=\linewidth,trim={0 30 0 45},clip]{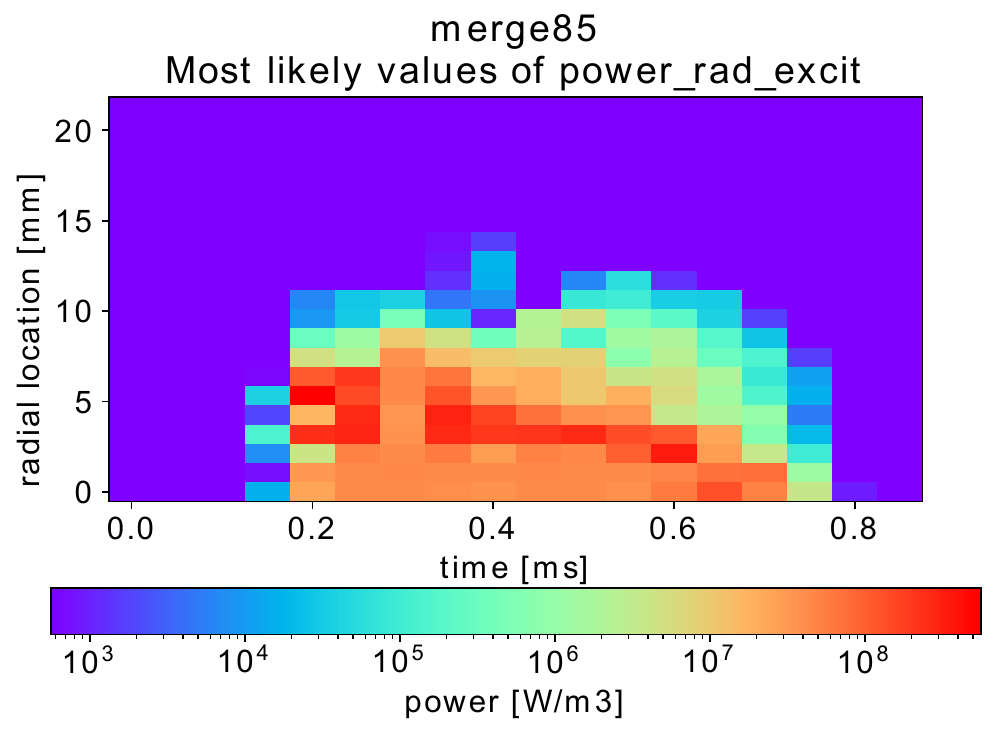}
         \vspace*{-5mm}
         \caption{radiated power EIE [$W/m^3$]}
        \label{fig:bayes_example_1a}
    \end{subfigure}
    \hfill
    \begin{subfigure}{0.8\linewidth}
    	\includegraphics[width=\linewidth,trim={0 30 0 45},clip]{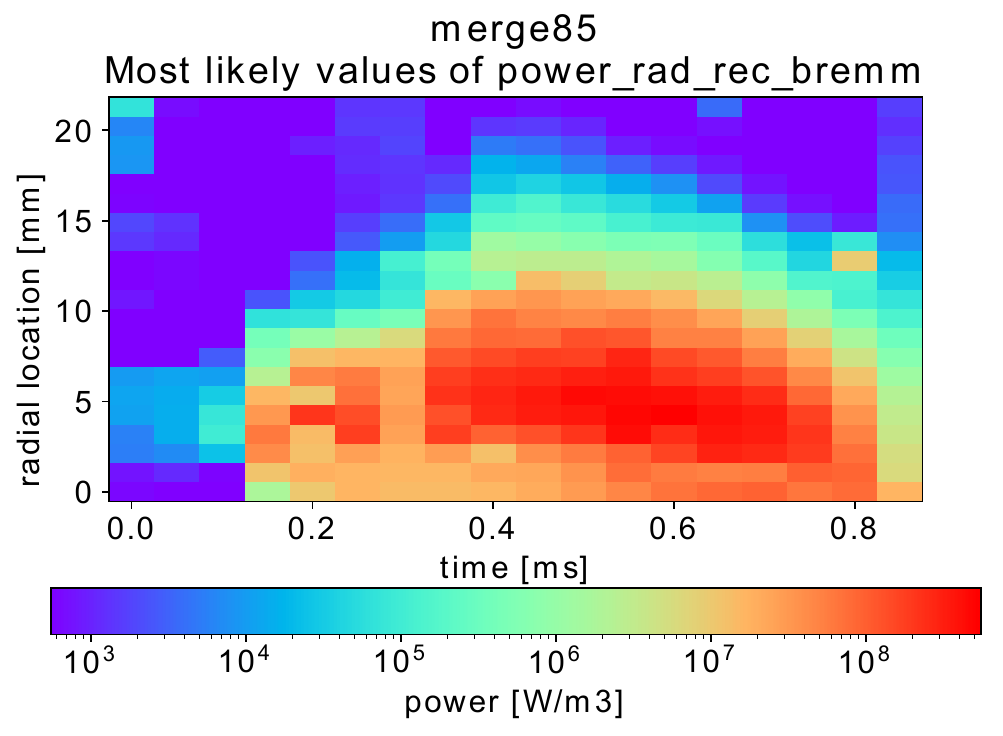}
         \vspace*{-5mm}
         \caption{radiated power recombination [$W/m^3$]}
        \label{fig:bayes_example_1b}
    \end{subfigure}
    \hfill
    \begin{subfigure}{0.8\linewidth}
    	\includegraphics[width=\linewidth,trim={0 30 0 45},clip]{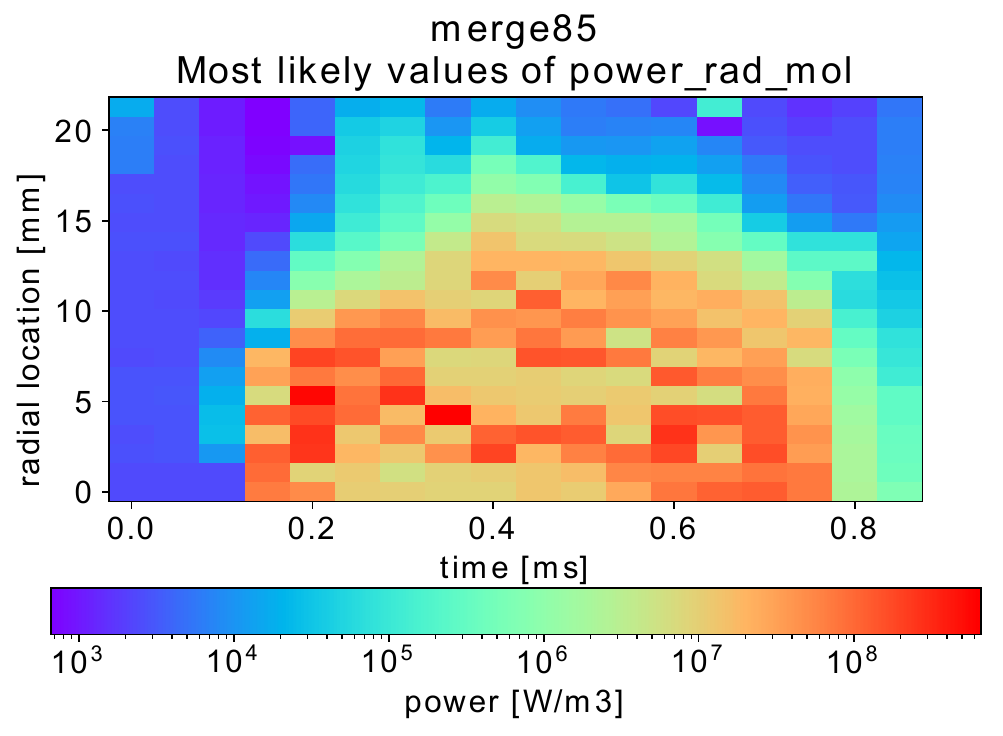}
         \vspace*{-5mm}
         \caption{radiated power MAR/MAI/MAD [$W/m^3$]}
        \label{fig:bayes_example_1c}
    \end{subfigure}
	\caption{Most likely values of the radiated power via EIE (\subref{fig:bayes_example_1a}), recombination (\subref{fig:bayes_example_1b}) and molecular processes (\subref{fig:bayes_example_1c}) for ID 10 in \autoref{tab:table1} (same conditions of shown in Figure 7b of \cite{Federici}). The noise shown in these figures arises from the fact that the analysis is independent for each radial and temporal location (no regularisation applied).}
    \label{fig:bayes_example_1}
\end{figure}

The results at each time and radial location, as explained in \autoref{Bayesian analysis}, consist of a collection of PDFs for the quantities of interest in the volume from target skimmer to target, like the power radiated via EIE. The temporal and spatial distribution of the most likely values can already be useful, as they can inform on the conditions in which some phenomena are important. Our inference shows that ionisation and excitation tend to be more important close to the axis of the plasma column, where and when temperature is highest, but are not peaked on the axis as the atomic hydrogen density profile is hollow. Radiation from EIR is strong for high density and intermediate temperature. Radiation from excited H atoms ($H^*$) generated from plasma molecular reactions (PMR) is present at similar times and locations as EIR but is significant up to even larger radii (and lower temperatures) than EIR. This is shown for ID 10 in \autoref{tab:table1} in \autoref{fig:bayes_example_1}.

\begin{figure}
	\centering
	\includegraphics[width=\linewidth,trim={0 332 0 0},clip]{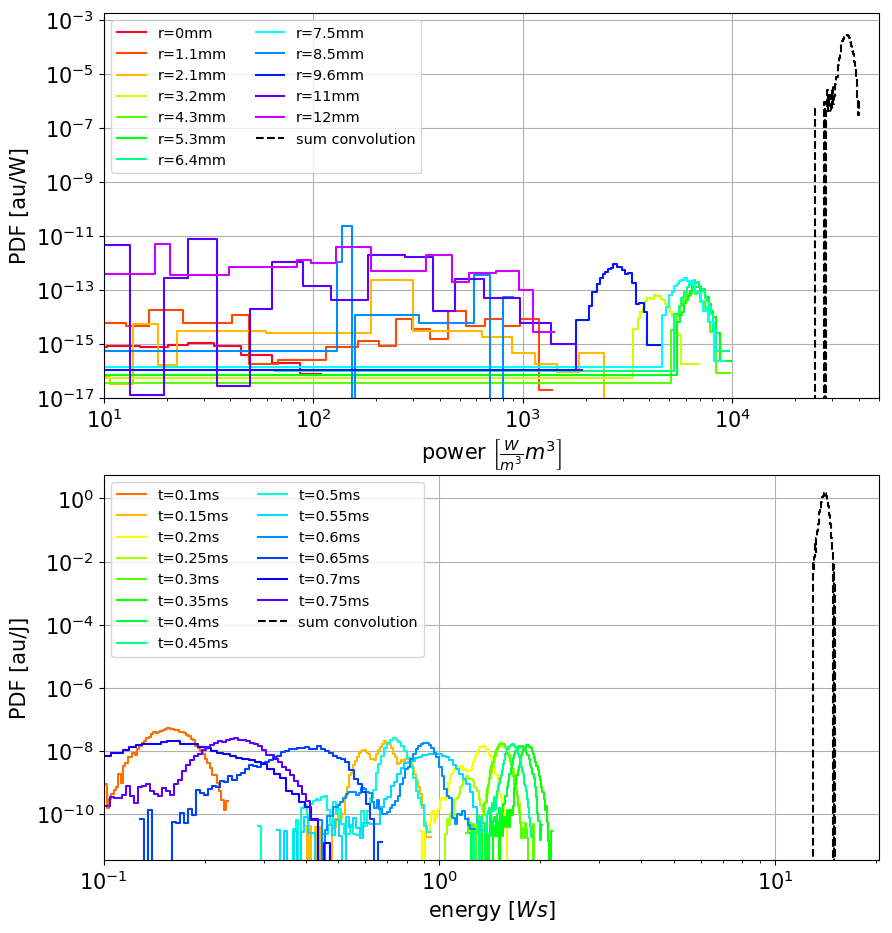}
	\caption{Example of the radial sum of the power radiated via excitation for ID 5 in \autoref{tab:table1} at 0.4ms. Only the radial locations that are more relevant are here shown. The power density ($W/m^3$) at each location is multiplied by the volume it pertains to determine the x coordinate.}
	\label{fig:bayes_example_r_sum}
\end{figure}

The PDFs are convolved over the radial direction to integrate the parameter of interest on the radii. An example of the radial convolution of the power radiated from EIE at 0.4ms for ID 5 in \autoref{tab:table1} is provided in \autoref{fig:bayes_example_r_sum}. It can be noted that some PDFs, for example at $r=11mm$, show multiple peaks. This indicates that multiple possible combinations of the precursors densities with similar likelihood have been found.

\begin{figure}
	\centering
     \begin{subfigure}{1\linewidth}
    	\includegraphics[width=\linewidth,trim={5 0 5 45},clip]{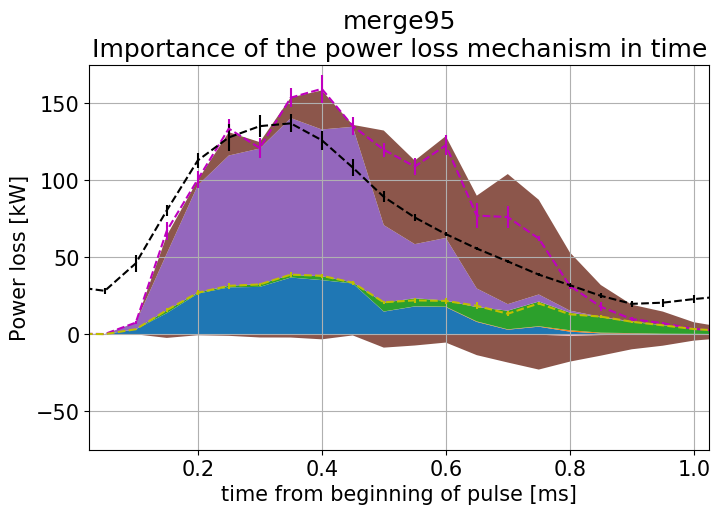}
         \vspace*{-5mm}
         \caption{neutral pressure = 0.3Pa}
        \label{fig:bayes_example_2a}
    \end{subfigure}
    \hfill
    \begin{subfigure}{1\linewidth}
    	\includegraphics[width=\linewidth,trim={5 0 5 45},clip]{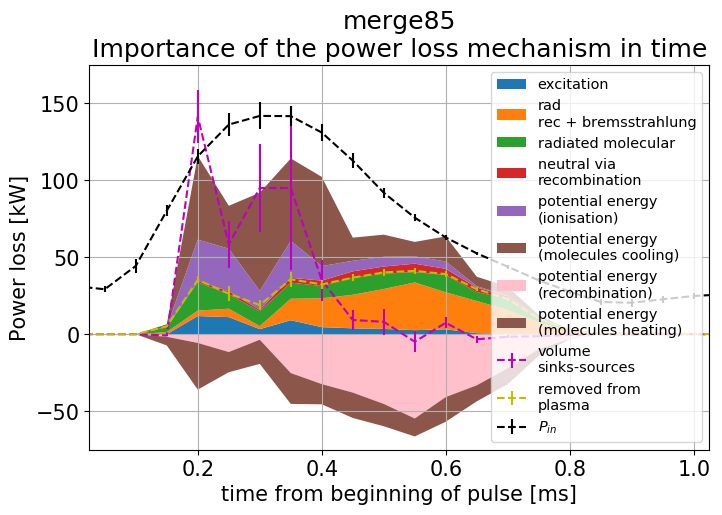}
         \vspace*{-5mm}
         \caption{neutral pressure = 15Pa}
        \label{fig:bayes_example_2b}
    \end{subfigure}
	\caption{Stacked area plot of the most likely values for the terms of the power balance as per \autoref{Power balance} for the lowest (\subref{fig:bayes_example_2a}) and highest (\subref{fig:bayes_example_2b}) pressure conditions for strong pulses. The data below zero corresponds to energy that released by recombination reactions heating the plasma. The potential energy from interactions with molecules is split in its positive and negative components. As \emph{molecules} is here intended any reaction with exchange of potential energy different to EIR and ionisation (see \autoref{Reactions}). \emph{volume sinks-sources} corresponds to the net local power balance while \emph{removed from plasma} indicates the contribution to the plasma column power balance as per \autoref{Plasma column power balance}. The colors in the legend of (\subref{fig:bayes_example_2b}) apply to both plots.}
    \label{fig:bayes_example_2}
\end{figure}

The evolution over time of the terms of the local the power balance, as per \autoref{Power balance}, are shown in \autoref{fig:bayes_example_2} for the lowest and highest pressure cases.
In \autoref{fig:bayes_example_2a} is shown that the energy of the ELM-like pulse is almost entirely used for ionisation, dissociation and molecular processes (MAD, MAI) increasing the plasma potential energy. However these ions recombine at the target, meaning that a larger fraction of the pulse energy will be able to reach the target, and in fact for this condition the target receives the most heat. The net power sink from the plasma is slightly larger than the input energy after its peak. This can be understood as there is no mechanism to enforce a global power balance since the analysis for every radial and temporal location is independent. The radiated energy is mostly due to EIE, that is to be expected due to the high temperature. After 0.6ms the temperature decreases below 4eV and the radiative and potential energy losses to the plasma from MAI/MAD and dissociation increases.

For higher neutral pressure (\autoref{fig:bayes_example_2b}) the peak in power input coincides to the peak of power loss from molecular reactions instead of ionisation, primarily due to the lower temperature. Most importantly, after the peak in input power and temperature, there is a strong influence of EIR. Radiative losses and potential energy gains balance such that after 0.4ms the net power sink on the plasma is negligible.\cite{Verhaegh2019} The period of peak radiative losses now coincides with the peak in EIR. More energy is radiated then before and recombination in the volume means that the plasma is depleted before reaching the target. The radiative losses are on the expenses of the plasma potential rather than its thermal energy. This causes a reduction of the heat flux just like in deep detachment, matching with the results from thermographic observations.\cite{Federici} EIR dominates the path to radiative losses but the influence of molecular reactions increases.
These results are quite similar to recent studies of detachment in medium size tokamaks including molecular precursors.\cite{Verhaegh2022,Verhaegh2021}

\begin{figure}
	\centering
	\includegraphics[width=\linewidth,trim={0 0 30 60},clip]{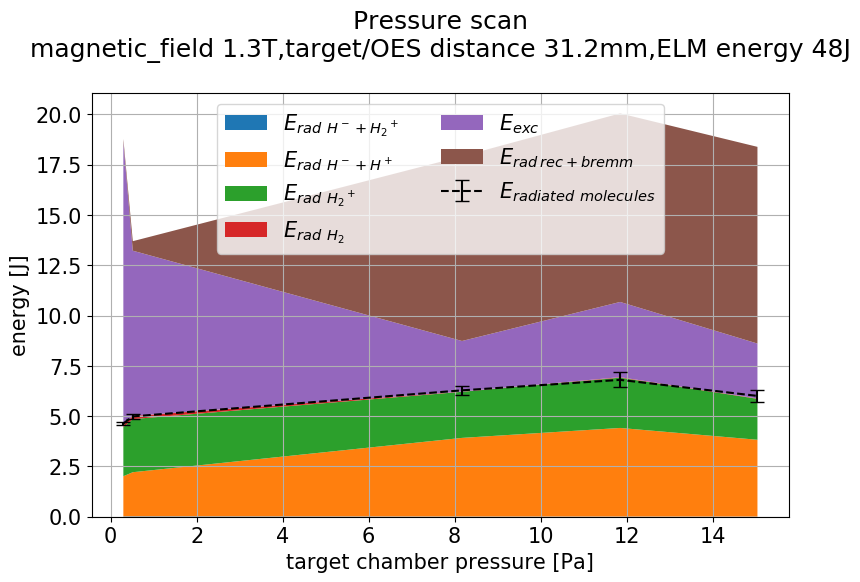}
	\caption{Energy dissipated radiatively per mechanism for increasing neutral pressure (ID 5 to 10 in \autoref{tab:table1}) obtained with the Bayesian analysis.}
	\label{fig:bayes_example_3}
\end{figure}

The PDFs at each time are further convolved multiplying by the time interval they refer to, returning the total for the quantity of interest over the entire duration of the ELM-like pulse. In \autoref{fig:bayes_example_3} is shown how the total energy radiated by the plasma changes with target chamber pressure.
As the target chamber pressure is increased (Stage 1 (P$<$2Pa) to 2 (P$>$2Pa)) the dominant process changes from EIE to EIR and the impact of molecules increases reaching up to $\sim$1/3 of the total radiated power. This is also reflected in the total radiated power in the visible wavelength range (not shown): in Stage 2 it is 3 times higher than in Stage 1, 0.3J and 0.1J respectively (the Lyman series is always responsible for most of the power radiated), largely due to the increase of EIR importance. This is not as a dramatic increase as observed from thermography (Figure 5 in \cite{Federici}), implying that the extrapolated $H\alpha$ emission, from our inferences, could be overestimated. This could imply an overestimation of molecular assisted reactions in Stage 1 (visible radiated power losses due to EIE are 0.005-0.01J in for ID 5 and 6 in \autoref{tab:table1} respectively, with a better agreement with thermography). An overestimation of $H\alpha$ can also arise from an (over)underestimation of ($H^-$)$H_2^+$, changing the $H\beta/H\alpha$ ratio as shown in \cite{Verhaegh2020}. Among radiation due to molecules the precursor that is more important is $H^-$. This would be against results from previous analysis that show a dominance of the ${H_2}^+$ precursor\cite{Akkermans2020} but, as mentioned in \autoref{Limitations due to the measured lines}, the radiation from ${H_2}^+$ dissociation and $H^+$ mutual neutralisation has similar line ratios for the lines used in this study, so cannot be distinguished with certainty.

\begin{figure}
    	\includegraphics[width=\linewidth,trim={0 0 0 5},clip]{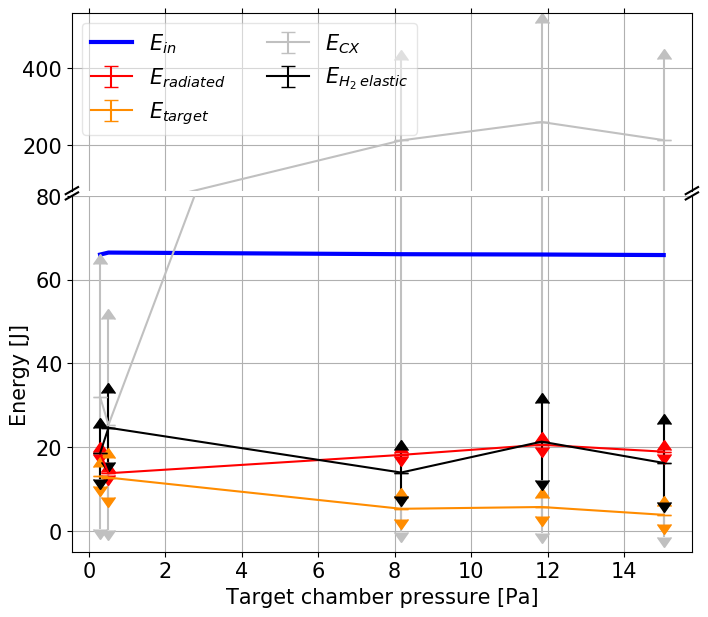}
	\caption{Results from the Bayesian analysis for strong ELM-like pulses with increasing neutral pressure (ID 5 to 10 in \autoref{tab:table1}). Total energy input (blue), energy removed from the plasma column (approximated with the hydrogenic radiative losses, red), energy absorbed by the target from thermographic analysis\cite{Federici} (orange), estimated maximum energy removed from CX (gray) and $H_2$ elastic (black) collisions. CX and $H_2$ elastic have already been multiplied by the corrective factors found in \autoref{Bayesian analysis} by comparison with B2.5-Eunomia.}
	\label{fig:bayes2}
\end{figure}

The variation of the terms in the global plasma column power balance with respect to target chamber pressure is presented in \autoref{fig:bayes2}. The energy absorbed by the target was determined from thermographic analysis.\cite{Federici} The estimation of CX and $H_2$ elastic collisions is as per \autoref{Power balance}. The two terms are already multiplied by the corrective factors found by comparing their simplified model with B2.5-Eunomia, identified in \autoref{Bayesian analysis} (CX and $H_2$ elastic collision losses from B2.5-Eunomia being 1-150\% and 18-34\% respectively compared to the simplified estimate). The radiative energy losses increase with target chamber neutral pressure, up to $\sim$1/3 of the input energy, while the energy to the target decreases. This transition corresponds to the transition between Stage 1 and 2 introduced when analysing the visible fast camera brightness. $H_2$ elastic losses are of the same order of magnitude as radiative and target losses, while CX has a very large range from zero to one order of magnitude higher than $P_{in}$. These losses are here only estimated. Subtracting the radiative losses from the input energy returns much more than the measured target flux, allowing for other loss channels like CX, $H_2$ elastic collisions or the losses associated with plasma surface interactions, that are here not inferred.

From this very crude balance it seems that losses due to radiation and energy transfer to the target can account for about 1/3 of the input energy and $H_2$ elastic collisions for another 1/3. From fast camera observations, for high pressure conditions the energy dissipated at the target could be less relevant, leaving more room for CX losses, while the opposite for low pressure.

\subsubsection{Particle balance}\label{Particle balance bayesian}
During the experiment there was no direct measurement of any part of the particle balance, so it is not as well constrained as the power balance. Nevertheless thanks to the assumptions and approximations in \autoref{Balance over the plasma column} it is possible to perform a rough particle balance on the plasma column. Only the charged particles' particle balance is utilised, as only these are confined to a certain poloidal location by the magnetic field. Only $H^+, e^-$ are introduced by the plasma source, using the flow velocity estimated from the power input in \autoref{Balance over the plasma column}. The volumetric particle sources and sinks are calculated together with the power ones and the target particle losses are treated as an unknown in a similar way as in the power balance. To calculate the prior probability, the balance of $H^+$ and $e^-$ has a large uncertainty, as the plasma can potentially accumulate in the plasma column, while the balance of ${H_2}^+$ and $H^-$ has a smaller uncertainty due to their short lifetime. More detail on the particle balance is given in \ref{Particle balance}.  

The procedure to convolve the data from time and spatial dependent PDFs to a global one for the whole pulse is the same as previously mentioned for the power. In order to better understand the influence of molecular reactions on the plasma the individual reaction rates are divided in MAR, MAI and MAD. These are defined as the paths composed of a first reaction that converts a neutral specie ($H_2$, $H$) into a molecular precursor (${H_2}^+$, $H^-$) then a second one where the precursor is used. The final result of the two reactions combined can either be the recombination of the plasma that participated to the reactions (MAR) the ionisation of the original neutral (MAI) or the dissociation of $H_2$ (MAD). Here paths that cause ionisation or recombination and dissociation are not accounted in MAD to avoid double counting. All of the possible combinations of reactions that achieve the mentioned goals via a molecular precursor are added together to form the MAR, MAI and MAD rates.\cite{Verhaegh2020} 

\begin{figure}
    	\includegraphics[width=\linewidth,trim={30 260 60 60},clip]{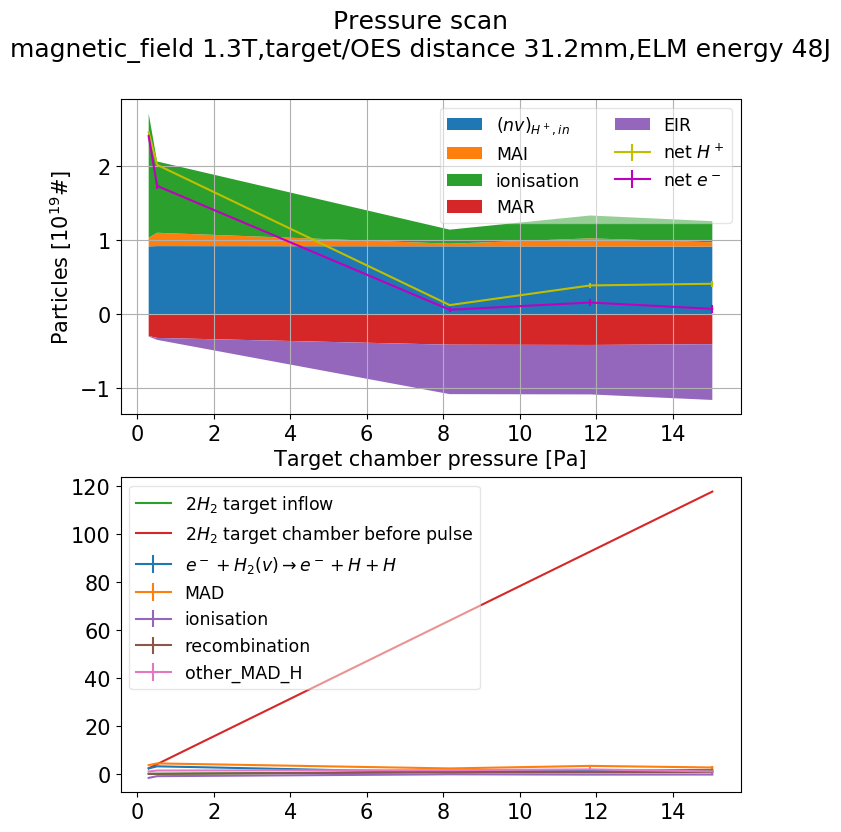}
	\caption{Plasma particle balance on the plasma column from the Bayesian analysis for strong ELM-like pulses with increasing neutral pressure (ID 5 to 10 in \autoref{tab:table1}). Total particle input from the plasma source estimated from \autoref{Balance over the plasma column} and \ref{Particle balance} ($(nv)_{H^+,in}$). Number of plasma particles created during the duration of the ELM-like pulse by type of reaction. MAR and MAI are defined as ensemble of multiple reactions that, starting from neutrals, first create a molecular precursor (${H_2}^+$, $H^-$) then further react with the end result of the two consecutive reactions being plasma recombination or ionisation respectively.\cite{Verhaegh2020} We note that there can be a discrepancy between net$H^+$ and net$e^-$, wich appears to be an analysis artefact.}
	\label{fig:bayes3}
\end{figure}

The particle balance on the plasma is shown in \autoref{fig:bayes3}.
Here are shown the plasma particle input, constituted by the source, the ionisation source and the recombination sink. MAI increases at lower pressure but is not very important representing a small fraction of the total ionisation. With the considered reactions ${H_2}^+$ is the only precursor that can cause MAI. MAR increases with pressure, but it dominates at low pressure (high temperature), and amounts to $\sim$1/3 of the total plasma particle sink at high pressure (low temperature). It is mainly due to the $H^-$ precursor. 

An important observation is that for low pressure conditions the ionisation source is higher than the estimated particle input from the source. The main difference between linear machines and tokamak divertors is that in a divertor, especially in high recycling, most of the particle source for the plasma comes from the recycling neutrals while the upstream plasma acts as the energy source. Conversely in linear machines the external source usually dominates. For low pressure conditions this does not happen here. If the neutrals predominantly come from plasma recycling at the target surface, this might be evidence of recycling in a similar fashion as in a tokamak. 
\autoref{fig:bayes3} shows a decrease in net balance of protons and electrons in the plasma column (net $H^+$ and net $e^-$) for increasing neutral pressure due to a transition from an ionising to a recombining plasma. This is in agreement with both the analysis of power losses as per \autoref{Power balance bayesian} and the target heating analysis
. The potential energy exchanges in the plasma like ionisation, dissociation, MAI and MAD are important to lower the plasma temperature enough for MAR and EIR to become important and reduce the particle flux to the target. MAD is more efficient than electron impact dissociation (EID) in converting $H_2$ to $H$ and MAR can effect the plasma at higher temperature than EIR, therefore the inclusion of the chemistry from ${H_2}^+$ and $H^-$ is important in describing the ELM burn through processes.

\begin{figure}
    	\includegraphics[width=\linewidth,trim={35 5 60 50},clip]{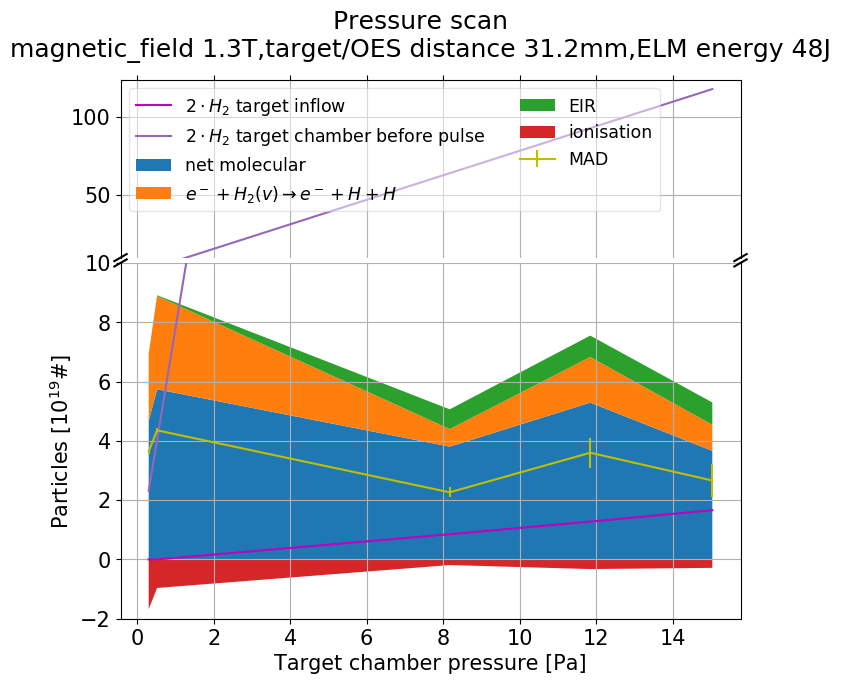}
	\caption{Atomic hydrogen particle balance on the plasma column from the Bayesian analysis for strong ELM-like pulses with increasing neutral pressure (ID 5 to 10 in \autoref{tab:table1}). It is assumed no atomic hydrogen enters the plasma column from the source thanks to differential pumping so only volumetric sources/sinks and interactions with the surface (not inferred here) are possible. Number of particles created during the duration of the ELM-like pulse by type of reaction. MAD is defined as ensemble of multiple reactions that, starting from neutrals, first create a molecular precursor (${H_2}^+$, $H^-$) then further react with the end result of the two consecutive reactions being exclusively $H_2$ dissociation.\cite{Verhaegh2020}}
	\label{fig:bayes4}
\end{figure}

The particle balance for atomic hydrogen is shown in \autoref{fig:bayes4}. It must be kept into account that this balance is not directly constrained and that transport is not considered, so the balance is very tentative. Hydrogen from molecular reactions, and MAD within it, represent the largest contribution to the source. Considering also that MAI is less then 10\% of MAD reaction rate, this means that most of the power losses due to molecules in the local power balance (\autoref{fig:bayes_example_2}) are due to MAD. Molecular reactions are responsible (integrating the local power balance) for about half of the potential energy exchange while up to a third of the radiative power losses. Molecular reactions seem to not have a strong impact in the plasma particle balance while dominating the neutral's particle balance. This is in agreement with previous findings in tokamaks such as TCV \cite{Verhaegh2021a}, JET \cite{Karhunen2023}, and TCV SOLPS-ITER modeling with reaction rates modified to properly assess molecular contribution \cite{Verhaegh2023}.

It must be noted that impact of molecular reactions on the particle balance, radiated power, and potential energy is heavily dependent on the rates used. For this work the Yacora rates have been used to determine the radiative power losses due to molecular precursors. To determine the reaction rates and the potential energy losses, other rates from AMJUEL and Janev have been used, with no guarantee of the consistency with Yacora (in terms of approximations used, consideration of different ro-vibrational states for $H_2$ and ${H_2}^+$ and other parameters used in the collisional radiative codes). For typical plasma conditions in this study ($T_e=2eV$, $n_e=10^{21}\#/m^3$), the average radiative power losses per molecular reaction are:
\begin{itemize}
    \item ${H_2}^+ + H^- \rightarrow H(p) + H_2$ : 0.17eV
    \item $H^+ + H^- \rightarrow H(p) + H$ : 8.2eV
    \item ${H_2}^+ + e^- \rightarrow H(p) + H(1)$ and \\ ${H_2}^+ + e^- \rightarrow H(p) + H^+ + e^-$ : 0.72eV
    \item $H_2 + e^- \rightarrow H(p) + H(1) + e$ : 0.026eV
\end{itemize}
The molecular rates are less established then atomic ones, so these numbers can change and significantly effect the power balance.


In low pressure conditions volume recombination is negligible and the vast majority of atomic hydrogen is produced by volumetric dissociation of $H_2$. Here the total volumetric source is higher then the total number of hydrogen atoms present in the target chamber before the ELM-like pulse and the neutrals that can be produced by surface recombination at the target (see \autoref{fig:bayes3}). This points to the source being over estimated, that can be expected being $H_2$ and $H$ particle balances not constrained. Both the hydrogen in the target chamber before the pulse and the possible neutrals generated at the target are larger then the volumetric ionisation sink. It is therefore unclear if what observed at low pressure is evidence of recycling or just dissociation and ionisation of a significant fraction of the gas in the target chamber.

A metric which is potentially significant for tokamaks is the average energy loss per interaction with the neutrals. The plasma enters the target chamber at about $10eV$, so Magnum-PSI can be thought as comparable to the section of a tokamak SOL from the region where the temperature the peak temperature is $\sim 10eV$ to the target. Above this temperature molecular effects become much less relevant, so the interactions between plasma and neutrals are easier to model. In Magnum-PSI, especially in Stage 2 where EIE is low, the importance of recycling appears to be very limited and the vast majority of the atomic hydrogen is produced via dissociation of $H_2$, that therefore can be considered the main neutral precursor causing the cooling of the plasma. The quantity of neutral gas that has a chance to interact with the plasma can be roughly estimated from the neutral pressure, assuming sonic inflow at room temperature over a cylinder or a conventional 2cm radius.
In Stage 2 the $H_2$ that reacts with the plasma is about a fifth of what enters from the sides. This is likely a product of the geometry of the plasma and elastic collisions heating the neutrals and causing dilution.\cite{DenHarder2015} The average power losses via radiation due to plasma interaction with $H_2$ (then $H$) neutrals (EIE and MAD/MAR/MAI), calculated as $(P_{rad} - P_{rad \; recomb})/(2\dot{n}_{H_2,sinks})$, correspond to $0.9-1.7eV/\#$ per interaction. Considering the overestimation of the MAD reaction rate mentioned before, this metric is likely slightly higher. Including the potential energy losses due to ionisation and molecular reactions the energy losses per nucleon are $\sim4.5eV$ in Stage 1 and $\sim6eV$ in Stage 2. Of these the fraction due to ionisation and MAI, that converts plasma thermal energy to potential and therefore requires recombination or MAR not to effect the target, is $~1eV$ and $2.5eV$ respectively. For comparison in Stage 2 the estimated energy losses due to CX and $H_2$ elastic collisions are $0.04-12eV$ and $0.15-0.5eV$ per nucleon entering the reference plasma volume respectively.

Another useful metric is the energy cost per ionisation\cite{Dudson2018,Verhaegh2021}: as temperature decreases and the plasma becomes more detached more excitation happens before ionisation occurs. In our case the radiation due to EIE increases from $\sim 5eV$ in Stage 1 to $\sim7eV$ in Stage 2, while it was $\sim22eV$ in TCV at particle flux roll-over.\cite{Verhaegh2019} The radiation due to molecular reactions per ionisation increases from $\sim2eV$ in Stage 1 to $\sim15eV$ in Stage 2, showing the relevance of molecular reactions and bringing the total closer to TCV estimations. Considering then that not all the hydrogen produced via dissociation is ionised, the total cost including the potential energy losses for molecular reactions and ionisation itself increase from $\sim35eV$ to $\sim60eV$.
It must be considered that while radiative losses are in part constrained by matching the line emission, an absolute measure, the potential ones are only inferred, and could change significantly if elastic collisions and CX are accounted for properly. Recycling, potentially significant in Stage 1, can cause an increase of the radiative losses via EIE at the target but also a local increase in plasma density, therefore only moderately reducing the heat flux to the target.

The results from particle balances are here tentative and obtained with strong approximations, so further investigations will be necessary to better understand the respective roles of the various sources and sinks.

\section{Summary}\label{Summary}

This study explores the impact of ELM-like pulses on a detached target in Magnum-PSI, employing a range of diagnostics. 
In the first paper associated to this study\cite{Federici}, it was observed that as the level of detachment increased, the energy of the ELM-like pulse was increasingly dissipated, eventually preventing it from reaching the target. This is attributed to increased volumetric power losses. The plasma's behaviour was classified into distinct stages. In Stage 1, the plasma remains attached to the target both before and during the ELM. By increasing the neutral pressure, the plasma id detached from the target prior to the ELM-like pulse but reattaches during the pulse, referred to as Stage 2. Detachment during the steady state is correlated with a significant rise in visible hydrogenic line emission, consistent with what is inferred from simulations.\cite{Zhou2022} Further increasing the neutral pressure results in the effective dissipation of the ELM-like pulse energy within the volume, and the plasma fails to reattach, marking Stage 3.
The power transferred to the target was independently calculated via thermography and demonstrated to decrease in accordance with the aforementioned division in stages. In the most extreme cases of Stage 3, target heating becomes undetectable, signifying the complete dissipation of the ELM-like pulse energy within the volume. This could also be applied to tokamaks by increasing the distance between the ionisation front and the target and increasing neutral compression, thus creating a cold gas buffer capable of dissipating a substantial portion of the ELM energy.


The optical emission spectrometer was upgraded to acquire temporally and poloidally resolved brightness in the visible range. Dedicated routines were developed to separate the spatial and temporal dependence in the raw data given by the rolling shutter, and an Abel inversion was performed on the brightness to give the line emissivity. Hydrogen Balmer line emission from OES ($p=4-8 \rightarrow 2$) was used to develop a Bayesian routine that incorporates the results from multiple diagnostics to return the power balance in the plasma column in the target chamber. Various approximations to extrapolate the results from a single location in the target chamber to its entirety were adopted. The routine incorporates priors from numerical simulations (B2.5-Eunomia) and collisional radiative codes (Yacora, ADAS). The goal is to understand what drives the previously identified volumetric losses and what the role of atomic and molecular assisted reactions is.

It was found that radiation from the plasma due to molecular assisted reactions is an important, although not dominant, energy loss mechanism. Radiation from excited hydrogen atoms created from plasma molecular interactions is found to be the about the same for Stage 1 and 2, but this is likely due to overestimation in Stage 1. Mutual neutralisation of ${H}^-$ seems to dominate radiative losses, but it was established that this could not be determined from OES alone with the present setup. Molecular assisted reactions significantly effect the local plasma power balance exchanging kinetic to potential energy, limiting what is available for ionisation and recycling. This is somewhat analogous to tokamaks, where power limitation of the ionisation source can induce detachment\cite{Verhaegh2018}.

Molecular reactions predominantly lead to MAD, as it is a more efficient path to $H_2$ dissociation than EID. These losses could be overestimated here due to an unconstrained particle balance for $H_2$ and $H$. Molecular processes are mostly dominant in an intermediate temperature range around 3eV, between ionisation and EIR dominant regions. The radiated power losses can be responsible for significant plasma power losses, but exchanges of potential energy like ionisation, EID, MAI, MAD, and collisional processes like CX and $H_2$ $H^+$ elastic collisions are expected to dominate the plasma power balance. Ionisation is more important at high temperature, $>$5eV, whereas at lower temperatures MAD/MAR increases significantly. When these processes cause the temperature of the plasma to go below $\sim$1.5eV EIR occurs, reducing significantly the plasma flow to the target and the relative heat flux. The significance of radiated power losses could differ if impurities such as carbon or nitrogen are present. These results indicate that for highly detached regimes in linear machines molecular interactions are important and need to be accounted, something that is not yet fully done in many codes used for Tokamak edge plasma studies.

\begin{acknowledgments}

The authors would like to thank: 
D. Wünderlich for his help in obtaining the Yacora coefficients; 
H. J. van der Meiden and J. Scholten for providing the Thomson scattering and ADC data.

This work is supported by US Department of Energy awards DE-AC05-00OR22725 and
DESC0014264 and under the auspices of the Engineering and Physical Sciences Research Council [EP/L01663X/1 and EP/W006839/1]. To obtain further information on the data and models underlying this paper please contact PublicationsManager@ukaea.uk.

Support for M. L. Reinke’s contributions was in part provided by Commonwealth Fusion Systems.

This work has been carried out within the framework of the EUROfusion Consortium, funded by the European Union via the Euratom Research and Training Programme (Grant Agreement No 101052200-EUROfusion). Views and opinions expressed are, however, those of the author(s) only and do not necessarily reflect those of the European Union or the European Commission. Neither the European Union nor the European Commission can be held responsible for them.

This work was supported in part by the DIFFER institute.

\end{acknowledgments}

\section{References}
\bibliographystyle{IEEEtran}
\bibliography{references}

\appendix

\section{OES data interpretation}\label{OES data interpretation}

\begin{figure}
	\centering
	\includegraphics[width=\linewidth,trim={440 50 600 150},clip]{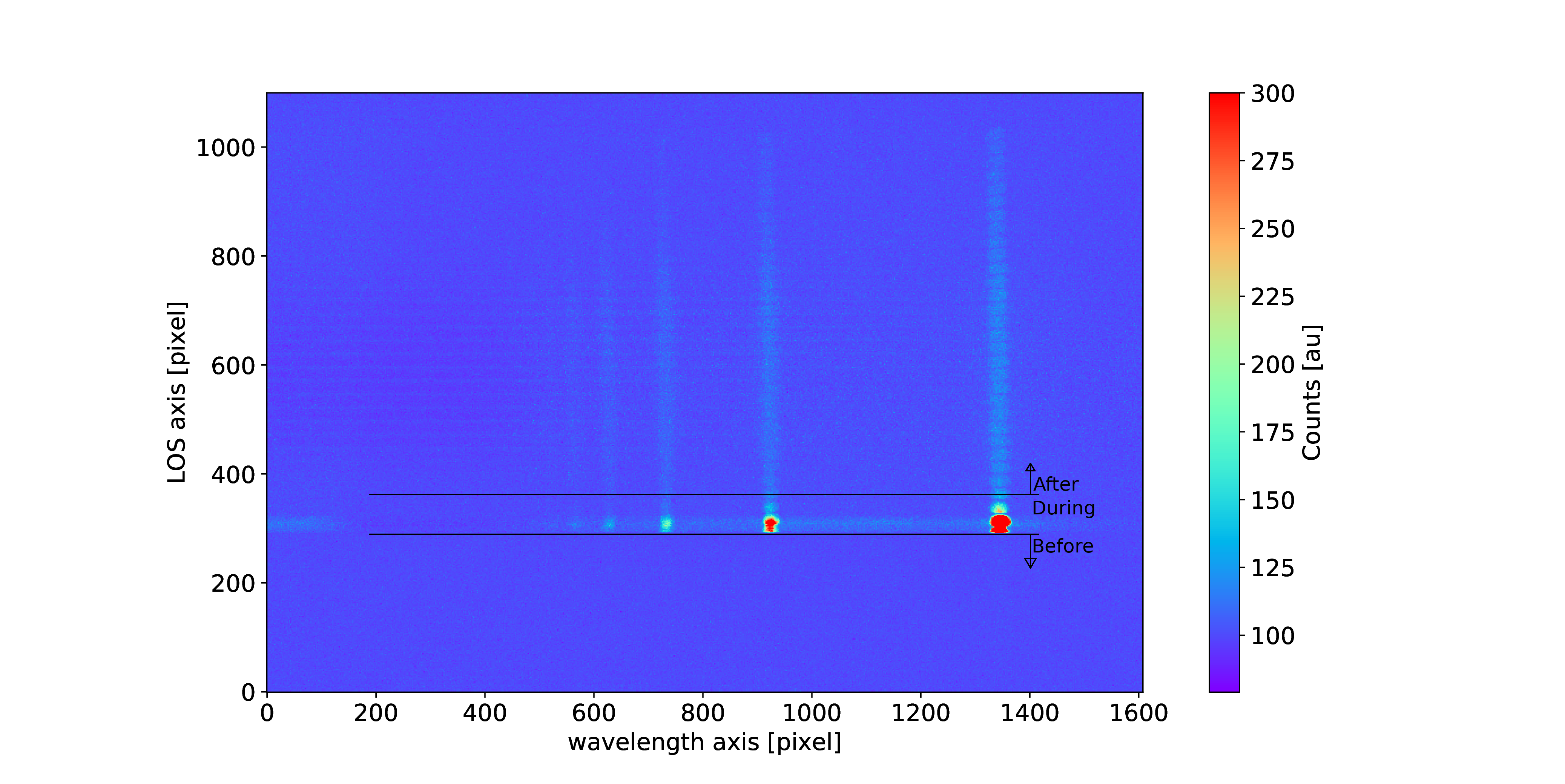}
	\caption{Example of a raw image from the OES. The readout starts from the bottom so higher rows represent later times. In the figure are indicated rows/times corresponding to different stages of the ELM-like pulse}
	\label{fig:sampling2}
\end{figure}

In this section, it will be detailed how the OES measurements are processed to obtain the local radially and temporally resolved emissivity used to estimate the relevance of molecular processes.
The camera that was selected for the purpose of collecting time resolved OES data was a Photometrics Prime95B 25mm RM16C, because of the relatively high signal to noise in low light conditions and large size of the sensor. In \autoref{fig:sampling2} is shown the typical picture collected during an experiment when on top of a steady state plasma an ELM-like pulse is fired. The rows are read sequentially from the bottom, with a time shift equal to the integration time, minimum $20\mu s$. This means that for the particular example shown the rows indicated as \emph{Before} represent times before the effect of the ELM-like pulse propagated to the OES location. \emph{During} represents the pulse and \emph{After} is for times after the ELM-like pulse, characterised by homogeneous line emission from the hot gas filling the target chamber. From \autoref{fig:sampling2} it is also possible to distinguish part of the 40 line of sight that are available.

\begin{figure}
	\centering
	\includegraphics[width=\linewidth,trim={100 30 290 200},clip]{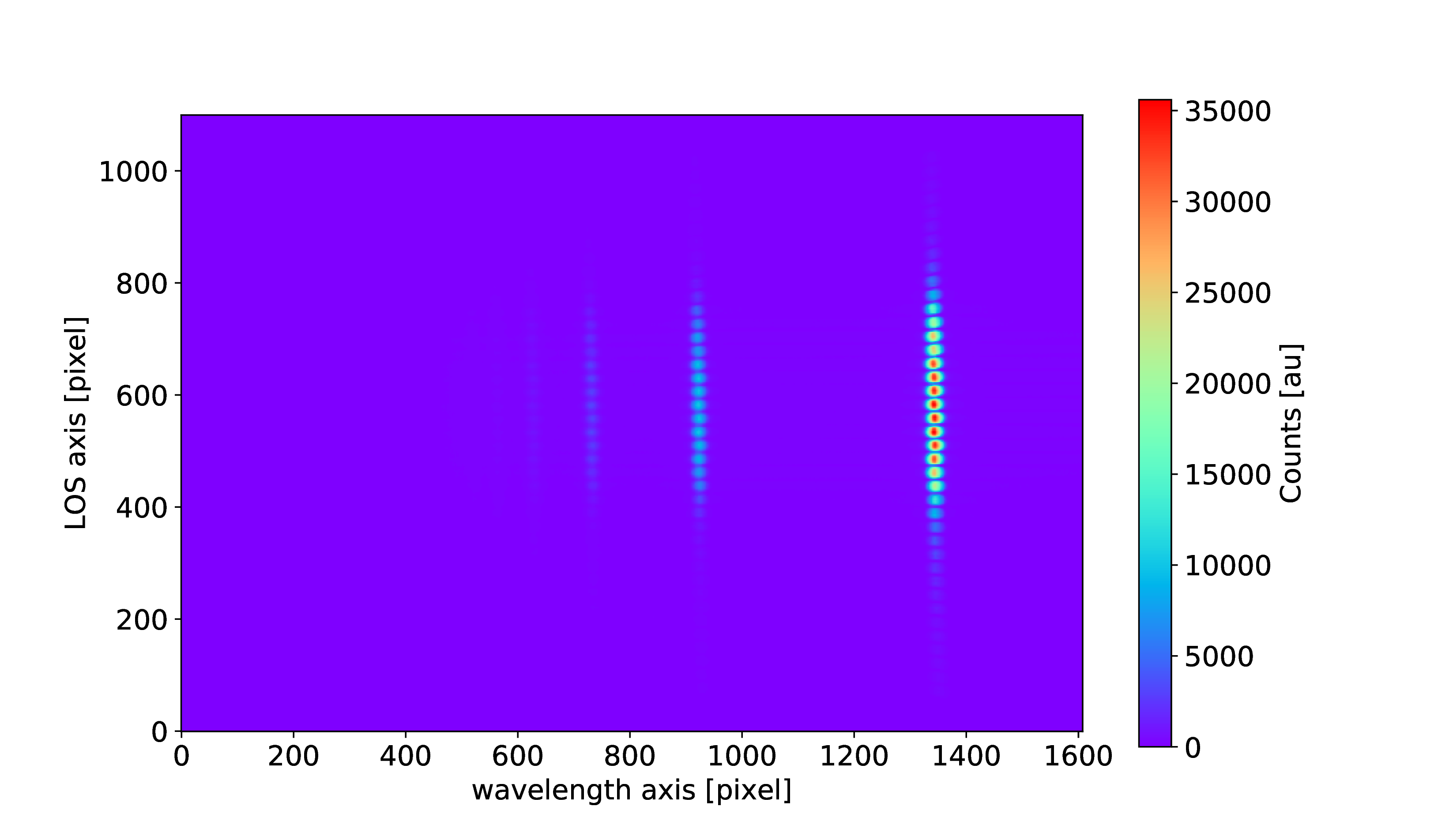}
	\caption{Example of decomposed time frame showing the symmetry of the image to the vertical pixel $\sim$600, likely representing the location of the plasma column axis.}
	\label{fig:sampling3}
\end{figure}

The first effect that is compensated is the sensitivity of the camera at low signal levels. The correlation between counts of the pixels and light intensity is mostly linear, but deviates significantly below 6 counts, and negative counts are returned at very low signal. A routine was developed to compensate for this, thanks to dedicated measurements to find the correlation between light intensity and counts.

To decouple spatial and temporal information, a scan is operated such that the ELM-like pulse is shifted in time with respect to the start of the camera image record and TS measurement. More details on the sampling strategy in Appendix A of \cite{Federici}. The presence of ELM-like pulses affected by capacitor bank misfires 
is found analysing the plasma source power and the data corresponding to those pulses is excluded. To separate the time and row dependency, for every row, column, and time of interest, the data in a range of $100\mu s$ and 8 rows is fit with a polynomial of second degree in time and first in row. To avoid over smoothing the image a smaller weight is assigned for increasing times and row difference from the one that is being examined. In \autoref{fig:sampling3} is shown the time/row decoupled image. The output time step has a $50\mu s$ resolution to match TS data.

\begin{figure}
     \centering
     \begin{subfigure}{0.8\linewidth}
         \includegraphics[width=\textwidth,trim={20 170 550 300},clip]{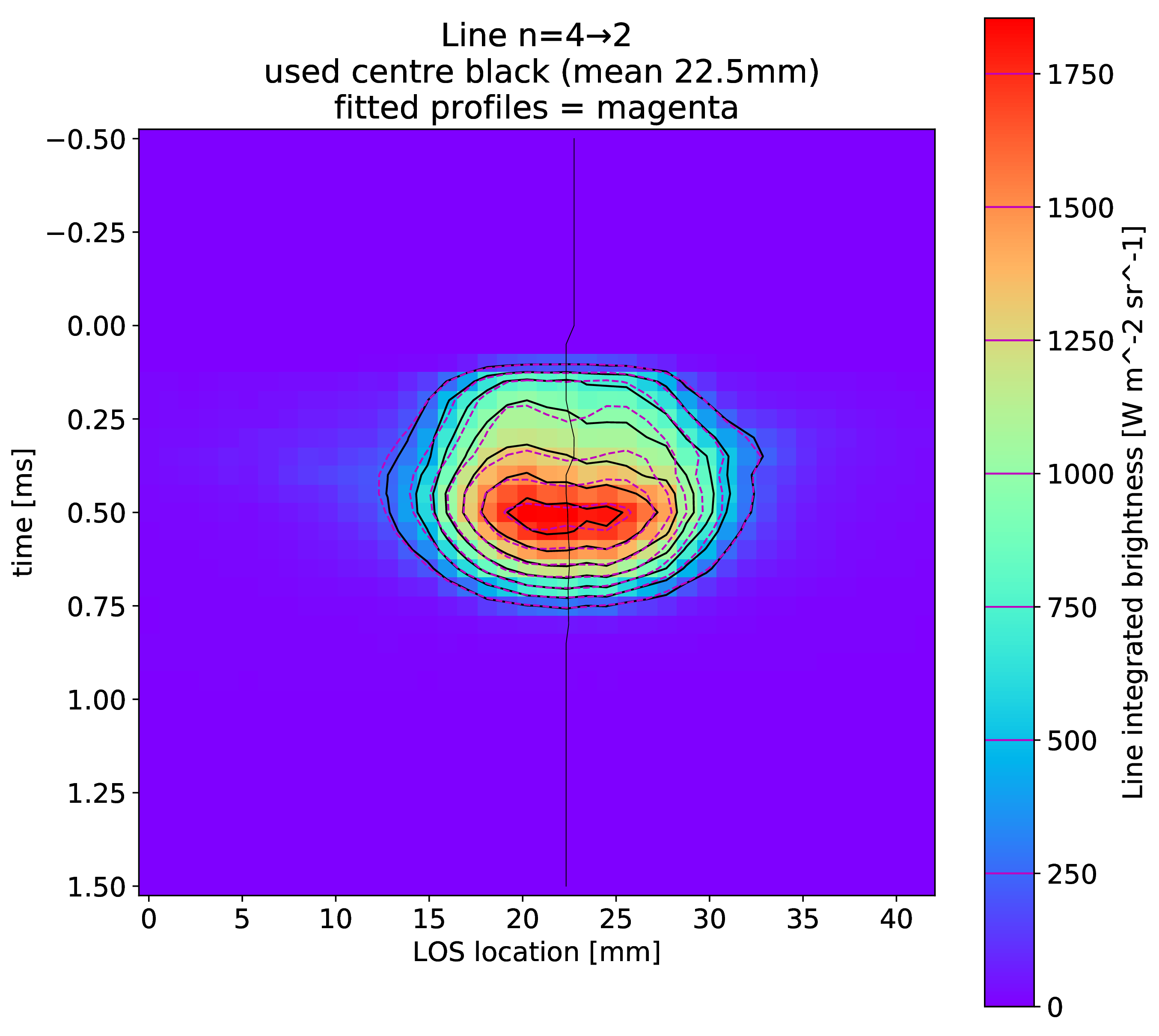}
         \caption{OES line integrated brightness}
         \label{fig:sampling4a}
     \end{subfigure}
     \begin{subfigure}{0.14\linewidth}
         \vspace*{-10mm}
         \includegraphics[width=\textwidth,trim={2200 0 0 40},clip]{chapter3/figs/line_integrted_profile4.png}
     \end{subfigure}
     \begin{subfigure}{0.8\linewidth}
         \includegraphics[width=\textwidth,trim={20 220 650 400},clip]{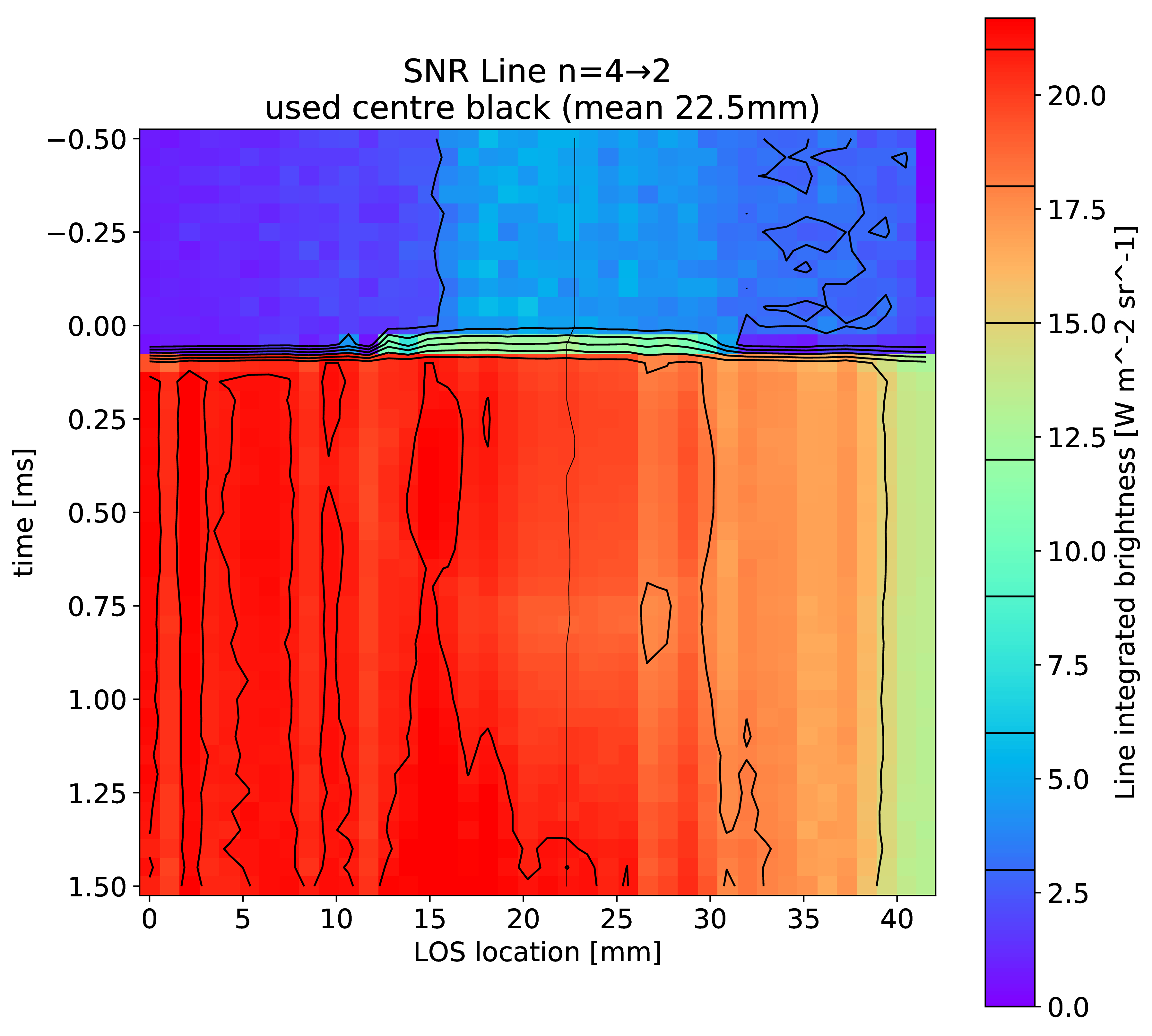}
         \caption{OES line integrated brightness SNR}
         \label{fig:sampling4b}
     \end{subfigure}
     \begin{subfigure}{0.095\linewidth}
         \vspace*{-10mm}
         \includegraphics[width=\textwidth,trim={3000 0 125 40},clip]{chapter3/figs/line4SNR.jpg}
     \end{subfigure}
        \caption{(\subref{fig:sampling4a}): example of a fit of the OES line integrated brightness and relative SNR with 3 Gaussians for the $n=4\rightarrow 2$ line. In magenta the profile of the fitted Gausinass, in thin black the found location of the plasma column centre. (\subref{fig:sampling4b}): SNR of the line integrated brightness.}
        \label{fig:sampling4}
\end{figure}

The counts are summed among the rows composing each LOS, and the line intensity is calculated by integrating above the background level. Brightness is then converted to emissivity via Abel inversion. The line emission is supposed to be poloidally symmetric and the plasma to be optically thin. In order to avoid unrealistic discontinuities given by noise, the superimposition of 3 Gaussian is fitted to the brightness profile as done by Barrois.\cite{Science2017} Each Gaussian can then analytically be Abel inverted and summed to obtain the total emissivity. In this process the uncertainties are propagated to be used in subsequent steps in the analysis. An example of the inversion process is shown in \autoref{fig:sampling4}. Given the signal to noise ratio and the available lines, it is decided to use Balmer lines $p=4-8 \rightarrow 2$.

\section{Details on the Bayesian calculations}\label{Details on the Bayesian calculations}

This section will provide a detailed explanation of how the expected properties of a plasma are calculated, compared with the measurements, and used in the particle and power balance, given a set of priors.

\subsection{Priors from B2.5 Eunomia}\label{Priors from B2.5 Eunomia}

To define the initial parameter space and the prior, it is necessary to define the range and probability associated with all the axis of the parameter space.

For $T_e$ and $n_e$, the TS values are used and the range is defined as 6 times the uncertainty. The probability is calculated as a linear normal distribution with the uncertainty corresponding to 1 sigma.

\begin{figure}
	\centering
	\includegraphics[width=\linewidth,trim={0 0 30 45},clip]{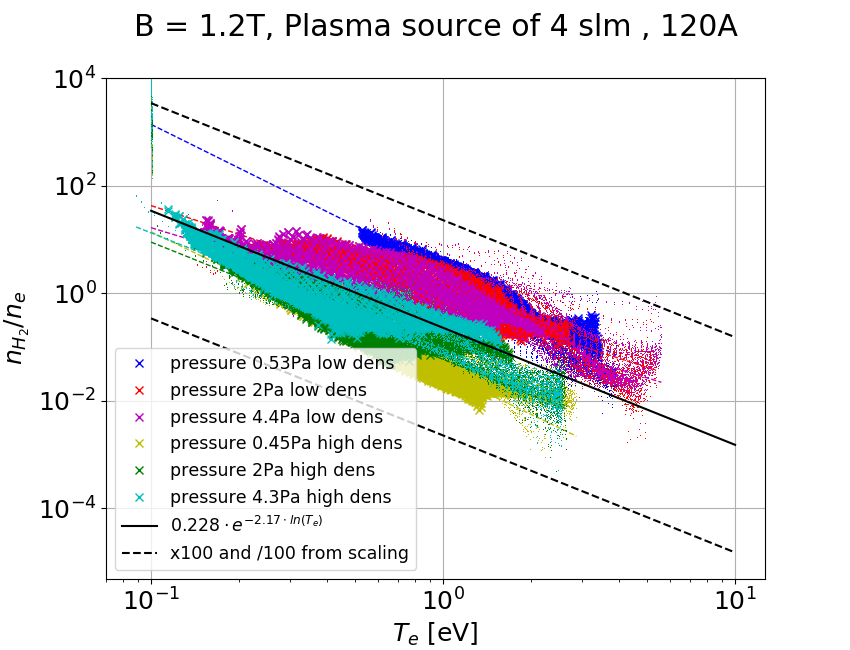}
	\caption{Correlation between molecular hydrogen and plasma density with temperature from B2.5-Eunomia modelling. The coloured dashed lines are obtained with a linear log log fit for the single cases. The solid black line is obtained averaging the fitting parameters obtained.}
	\label{fig:priors1}
\end{figure}

\begin{figure}
	\centering
	\includegraphics[width=\linewidth,trim={0 0 30 45},clip]{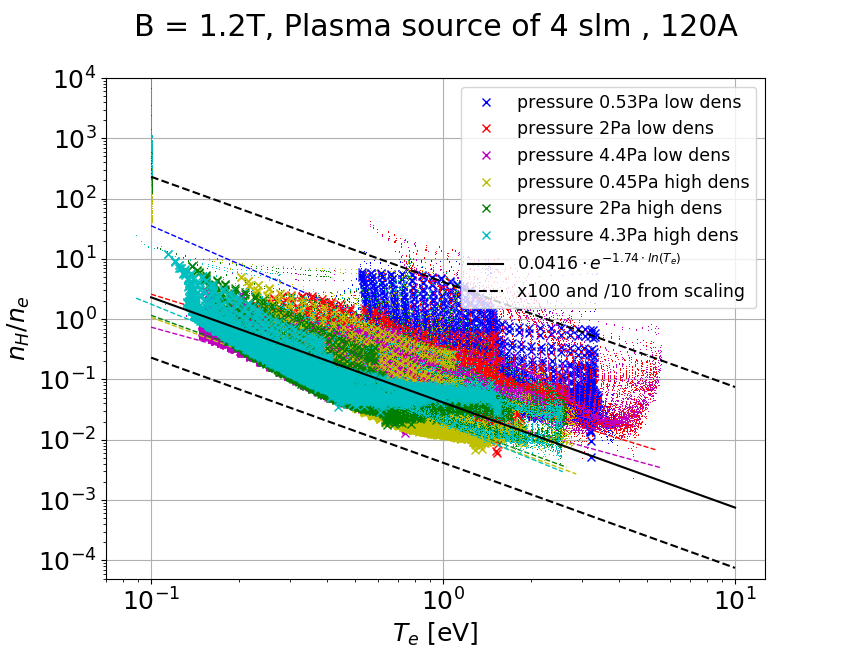}
	\caption{Correlation between atomic hydrogen and plasma density with temperature from B2.5-Eunomia modeling. The coloured dashed lines are obtained with a linear log log fit for the single cases. The solid black line is obtained averaging the fitting parameters obtained.}
	\label{fig:priors2}
\end{figure}

\begin{figure}
	\centering
	\includegraphics[width=\linewidth,trim={0 0 30 45},clip]{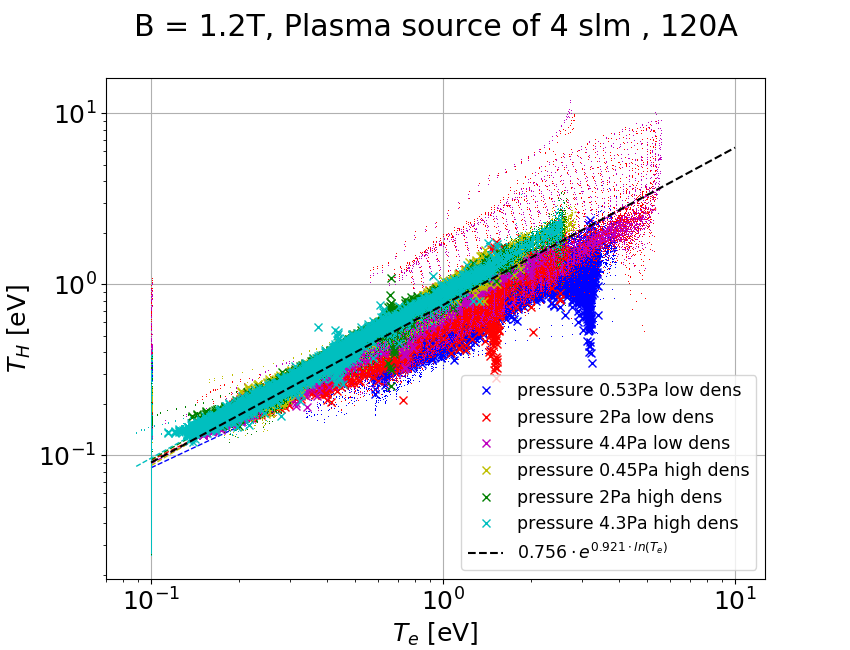}
	\caption{Correlation between atomic hydrogen and electron temperature from B2.5-Eunomia modeling}
	\label{fig:priors3}
\end{figure}

\begin{figure}
	\centering
	\includegraphics[width=\linewidth,trim={0 0 30 45},clip]{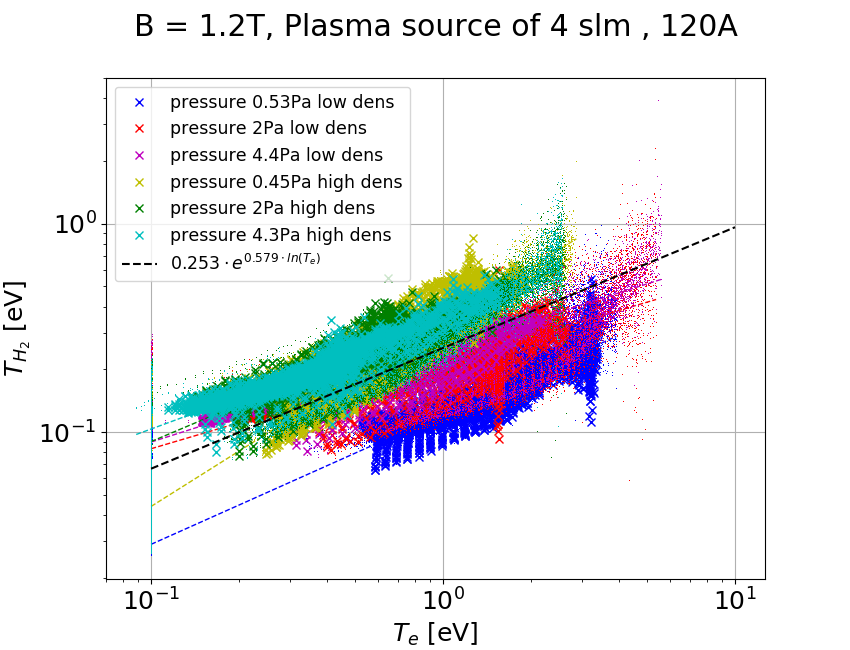}
	\caption{Correlation between $H_2$ and electron temperature from B2.5-Eunomia modelling. The coloured dashed lines are obtained with a linear log log fit for the single cases. The solid black line is obtained averaging the fitting parameters obtained. This is assumed to be the same as ${H_2}^+$ temperature, while to obtain $H^-$ temperature 2.2eV are added to account for the $H_2$ dissociation energy.}
	\label{fig:priors4}
\end{figure}

\begin{figure}
	\centering
	\includegraphics[width=\linewidth,trim={0 0 30 45},clip]{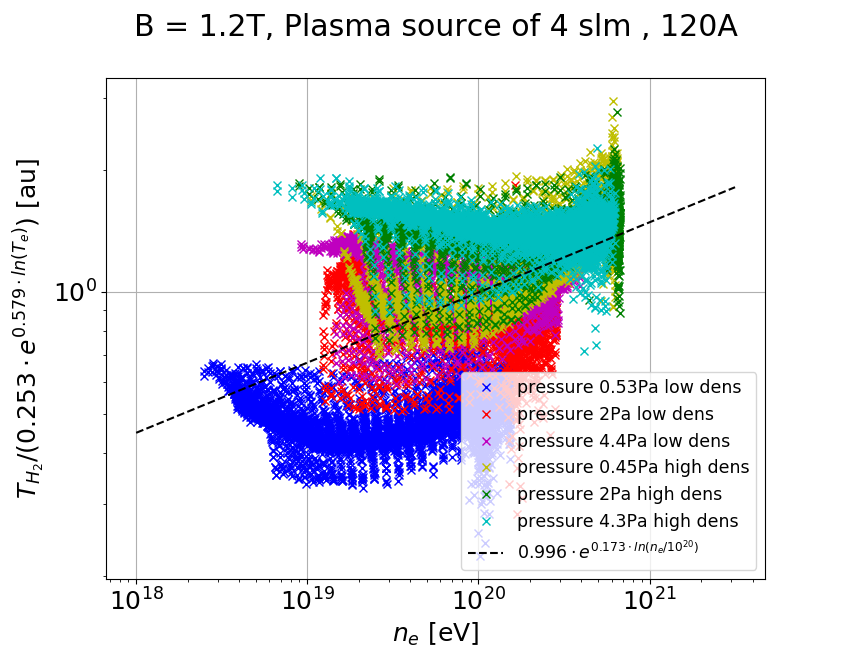}
	\caption{Residuals from fitting $T_{H_2}$ with the scaling from \autoref{fig:priors4} and their weak dependence on the plasma density. The linear log log scaling in black is obtained by fitting all points at once.}
	\label{fig:priors4b}
\end{figure}

\begin{figure}
	\centering
	\includegraphics[width=\linewidth,trim={0 0 30 45},clip]{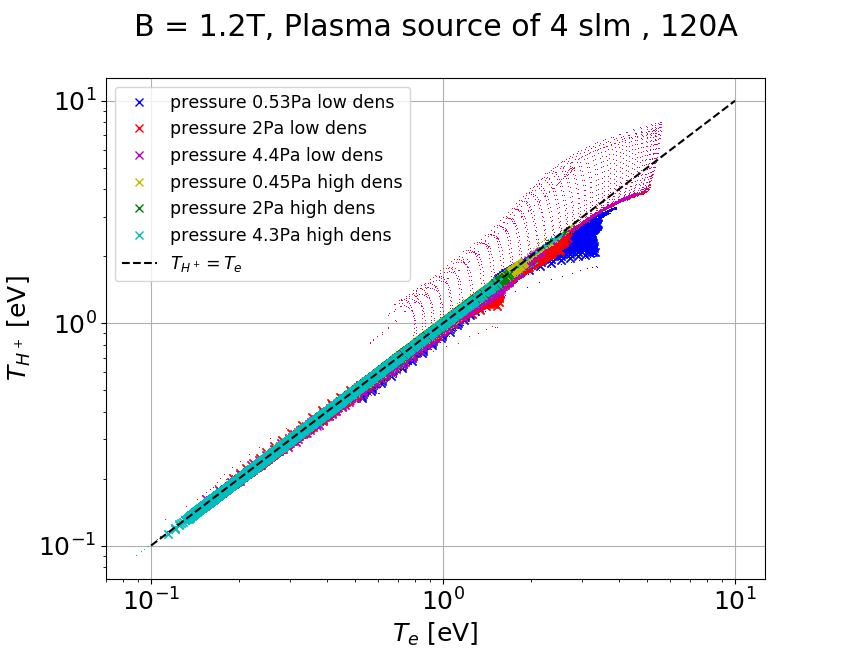}
	\caption{Correlation between $H^+$ and electron temperature from B2.5-Eunomia modelling. The dashed line indicates $T_{H^+}=T_e$.}
	\label{fig:priors5}
\end{figure}

The range and probabilities for $n_{H_2}/n_e$ are obtained from B2.5-Eunomia simulations for a steady state neutral pressure scan with 2 plasma source settings, ranging from attached to detached via increasing neutral hydrogen target chamber neutral pressure, carried out by Chandra.\cite{Chandra2021,Chandra2022} The simulations consider the whole plasma column source to target, but only data inside the target chamber and within 2cm of the axis are considered here (marked with $x$, as opposed to other regions marked by a point in \autoref{fig:priors1}, \ref{fig:priors2}, \ref{fig:priors3}, \ref{fig:priors4}, \ref{fig:priors4b}, \ref{fig:priors5}).

As demonstrated by Den Harder\cite{DenHarder2015}, the density decrease of $H_2$ in the plasma is mainly driven by rarefaction due to the high temperature of the plasma itself. For this reason a quite strong correlation between molecular hydrogen and plasma density ratio  and plasma temperature exists, shown in \autoref{fig:priors1}. The most likely ratio is obtained by fitting each simulation's results with a linear log log function and then averaging the fit parameters as shown in \autoref{fig:priors1}. The probability is defined as a normal distribution where 2 sigma corresponds to the black dashed lines, which are 100 and 1/100 times the fit value. The range is a significantly larger window around the dashed lines, to account for the large uncertainty coming from the fact that B2.5-Eunomia only simulates steady state conditions, while the ELM-like burn through is a very dynamic one.

The simulations are used to provide also range and probability for $n_H/n_e$. Atomic hydrogen is generated from recombination and from $H_2$ interaction with plasma and various molecules, so its density is only weakly correlated with plasma temperature and density, as shown in \autoref{fig:priors2}. 
The probability was calculated with a linear normal distribution with nominal value from the fit (calculated in the same fashion as for $n_{H_2}/n_e$) and 2 sigma arbitrarily assigned as per the dashed line in \autoref{fig:priors2}.

Other quantities that are part of the plasma state and had to be determined to calculate reaction rates and other coefficients are the temperatures of all species. To reduce the number of variables in the Bayesian algorithm their uncertainty is in this work not considered and only the nominal values are used. The correlations are shown in \autoref{fig:priors3}, \ref{fig:priors4}, \ref{fig:priors5} for $H$, $H_2$ and $H^+$ temperature respectively, where the black dashed lines indicates the values used. For $T_H$ and $T_{H_2}$ the fit is obtained in the same fashion as $n_H/n_e$ while for $T_{H^+}$ it is assumed $T_{H^+}=T_e=T_{plasma}$. For $T_{H_2}$ in particular a weak dependence on the plasma density is present, whose estimate is shown in \autoref{fig:priors4b} and can be due to an increase of the collisionality for higher density and a better coupling with the neutrals, resulting in a correction factor to apply to the dependency on $T_e$ alone. Given ${H_2}^+$ is mostly originated from $H_2$, it is considered $T_{H_2}=T_{{H_2}^+}$. This is valid for $H^-$ too, but because it can get some of the $H_2$ binding energy (2.2eV per atom), 2.2eV are added to $T_{H_2}$ to estimate $T_{H^-}$.\cite{Verhaegh2020}

\subsection{Priors from AMJUEL}\label{Priors from AMJUEL}
Ionised hydrogen molecules are generated mainly from $H_2$, so their density prior is calculated with AMJUEL\cite{Reiter2017}, a library that, among others, contains $n_{H^-}/n_{H_2}$ and $n_{{H_2}^+}/n_{H_2}$ density ratios in an equilibrium plasma for given plasma temperature and density (Section 12.58, 12.59, 11.11, 11.12). The conditions of an ELM-like pulse can potentially deviate significantly from equilibrium, so a wide range around equilibrium is considered as prior and a linear uniform distribution as likelihood.

\subsection{Priors range optimization}\label{Priors range optimization}
To optimize the $n_{H}/n_e$ range to only useful values, a combination of the information from \autoref{fig:priors1} and the OES measurement is used. For each $T_e$ / $n_e$ combination, the emission from EIR via the ADAS PEC coefficients is calculated and subtracted from the OES measurement. Then, the $n_{H}/n_e$ required to recreate each residual line emission via EIE plus its uncertainty is calculated. The largest $n_{H}/n_e$, limited by a predefined multiplier times the value from the B2.5-Eunomia fit, will be the highest value considered for that particular $T_e$ / $n_e$ combination. The lower limit will be taken as a small fraction of the maximum value, again limited by a predefined multiplier times the value from the B2.5-Eunomia fit. This way, parts of the range of $n_{H}/n_e$ that would cause an excessive line emission are automatically excluded and the prior range is assigned efficiently.

The same process is applied to the $n_{H_2}/n_e$ prior. For each $T_e$ / $n_e$ / $n_{H}/n_e$ combination, the total emission from EIR and EIE is subtracted from the OES measurement, and the $n_{H_2}/n_e$ required to match the residual is calculated using the Yacora coefficients for the $H_2$ dissociation reaction. For the $n_{{H_2}^+}/n_{H_2}$ prior, the emission from EIR, EIE, $H_2$ dissociation is considered. Consequently, for the $n_{{H}^-}/n_{H_2}$ prior the emission from ${H_2}^+$ is also considered.

\subsection{Emissivity}\label{Emissivity}

The emissivity is calculated for known precursors densities via the ADAS PECs and Yacora population coefficients.\cite{Verhaegh2020}
The Photon Emission Coefficients (PEC, photons $m^3/s$) coefficients are defined as the number of photons generated per second per unit of the precursors density. The number of photons for the transition $p \rightarrow q$ is equal to the product of density of the excited state $p$ and the Einstein coefficient $A_{pq}$. therefore, the emission generated by atomic excitation and recombination can be expressed as per \autoref{eq:emiss1} and \ref{eq:emiss2}, where what is intended as population coefficient ($PC_i$) is also highlighted.

\begin{equation}
\label{eq:emiss1}
\begin{aligned}
\epsilon^{exc}_{pq} = PEC^{exc}_{pq}(T_e,n_e) n_e n_{H} = A_{pq} \underbrace{ \frac{n_{H(p)}}{n_e n_{H}}}_{PC_{exc}} n_e n_{H} 
\end{aligned}
\end{equation}

\begin{equation}
\label{eq:emiss2}
\begin{aligned}
\epsilon^{rec}_{pq} = PEC^{rec}_{pq}(T_e,n_e) n_e n_{H^+{}} = A_{pq} \underbrace{\frac{n_{H(p)}}{n_e n_{H^+{}}}}_{PC_{rec}} n_e n_{H^+{}}
\end{aligned}
\end{equation}

The line emission due to molecular reactions is similarly calculated using the Yacora population coefficients as per \autoref{eq:emiss3}, \ref{eq:emiss4}, \ref{eq:emiss5} and \ref{eq:emiss6}. It is also shown which reaction was considered in building the coefficients, and the variables necessary to calculate the coefficients. As mentioned $T_H$, $T_{H_2}$ are determined from the B2.5-Eunomia simulation while $T_{H_2} \approx T_{{H_2}^+} \approx T_{H^-}-2.2eV$.

\begin{equation}
\label{eq:emiss3}
\begin{aligned}
\epsilon^{{H_2}^+{}}_{pq} =& A_{pq} PC_{{H_2}^+{}}(T_e,n_e) n_e n_{{H_2}^+{}} \\
reactions:\ &{H_2}^+{} + e^-{} \rightarrow H(p) + H^+{} + e^-{} \\ 
\ &{H_2}^+{} + e^-{} \rightarrow H(p) + H(0)
\end{aligned}
\end{equation}

\begin{equation}
\label{eq:emiss4}
\begin{aligned}
\epsilon^{{H_2}}_{pq} =& A_{pq} PC_{{H_2}}(T_e,n_e) n_e n_{{H_2}} \\
reaction:\ &{H_2}^+{} + e^-{} \rightarrow H(p) + H(1) + e^-{}
\end{aligned}
\end{equation}

\begin{equation}
\label{eq:emiss5}
\begin{aligned}
\epsilon^{{H}^-{}+{H_2}^+{}}_{pq} =& A_{pq} PC_{{H}^-{}+{H_2}^+{}}(T_e,T_{{H_2}^+{}},T_{{H}^-{}},n_e) n_{{H_2}^+{}} n_{{H}^-{}} \\
reaction:\ &{H}^-{}+{H_2}^+{} \rightarrow H(p) + H_2
\end{aligned}
\end{equation}

\begin{equation}
\label{eq:emiss6}
\begin{aligned}
\epsilon^{{H}^-{}+{H}^+{}}_{pq} =& A_{pq} PC_{{H}^-{}+{H}^+{}}(T_e,T_{{H}^+{}},T_{{H}^-{}},n_e) n_{{H}^+{}} n_{{H}^-{}} \\
reaction:\ &{H}^-{}+{H}^+{} \rightarrow H(p) + H(1)
\end{aligned}
\end{equation}

The total calculated emissivity and its uncertainty are determined as per \autoref{eq:emiss7} with $\sigma_{ADAS}$ and $\sigma_{Yacora}$ the uncertainty on the coefficients mentioned before.

\begin{equation}
\label{eq:emiss7}
\begin{aligned}
\epsilon^{calc}_{pq} =& \epsilon^{exc}_{pq} + \epsilon^{rec}_{pq} + \epsilon^{{H_2}^+{}}_{pq} + \epsilon^{{H_2}}_{pq} + \epsilon^{{H}^-{}+{H_2}^+{}}_{pq} + \epsilon^{{H}^-{}+{H}^+{}}_{pq}
\\
\sigma^{calc}_{\epsilon_{pq}} =& \left\{ {\sigma_{ADAS}}^2 \left({\epsilon^{exc}_{pq}}^2 + {\epsilon^{rec}_{pq}}^2\right) + \right. \\ &\left. + {\sigma_{Yacora}}^2 \left({\epsilon^{{H_2}^+{}}_{pq}}^2 + {\epsilon^{{H_2}}_{pq}}^2 + {\epsilon^{{H}^-{}+{H_2}^+{}}_{pq}}^2 + {\epsilon^{{H}^-{}+{H}^+{}}_{pq}}^2\right) \right\}^{1/2}
\end{aligned}
\end{equation}

The line emissivity measurement is compared with the expectation. For each precursor combination and line, the likelihood that $y_{pq}=0$ is calculated with \autoref{eq:emiss8a}
\begin{equation}
\label{eq:emiss8a}
\begin{aligned}
y_{pq} = \epsilon^{calc}_{pq}-\epsilon^{measure}_{pq} ,& \sigma_{y_{pq}} = \sqrt{{\sigma_{\epsilon_{pq}}^{calc}}^2 + {\sigma_{\epsilon_{pq}}^{measure}}^2}
\\
L(y_{pq} = 0|\Theta) =& \frac{1}{\sigma_{y_{pq}} \sqrt{2\pi}} e^{-\frac{1}{2} \left( \frac{y_{pq}}{\sigma_{y_{pq}}} \right)^2 }
\end{aligned}
\end{equation}
where $\Theta$ represent the specific combination of precursors that lead to the emission $\epsilon^{calc}_{pq}$.

Following Bayes theorem the posterior (probability of the combination of precursors generating the measurements) is calculated as the likelihood of the measurement being generated by the precursors times the probability associated with the precursors themselves divided by the probability of the measured data. For the case in which only one emission line is included in the model this is expressed in \autoref{eq:emiss8b}

\begin{equation}
\label{eq:emiss8b}
\begin{aligned}
P(\Theta|y_{pq} = 0) = \frac{L(y_{pq} = 0|\Theta) P(\Theta)}{P(y_{pq} = 0)}
\end{aligned}
\end{equation}

Where $P(y_{pq} = 0)$ acts as a normalisation factor. The final product of all probabilities will be normalised, so this term can be neglected. $P(\Theta)$ is  the product of the probability associated with every combination of precursors (see \autoref{Priors from B2.5 Eunomia}, \ref{Priors from AMJUEL} and \ref{Prior probability distribution}). The probability of fitting all the lines $P_{\epsilon}$ is then determined with \autoref{eq:emiss8}.

\begin{equation}
\label{eq:emiss8}
\begin{aligned}
P_{\epsilon} =& P(\Theta) \prod_{p=4,q=2}^{p=8} P(\Theta|y_{pq} = 0)
\end{aligned}
\end{equation}

For this calculations $\sigma_{ADAS}$ and $\sigma_{Yacora}$ were assumed 10\% and 20\%, respectively.

\subsection{Balance over the plasma column}\label{Balance over the plasma column}

\begin{figure*}
	\centering
	\includegraphics[width=0.7\linewidth,trim={0 0 0 0},clip]{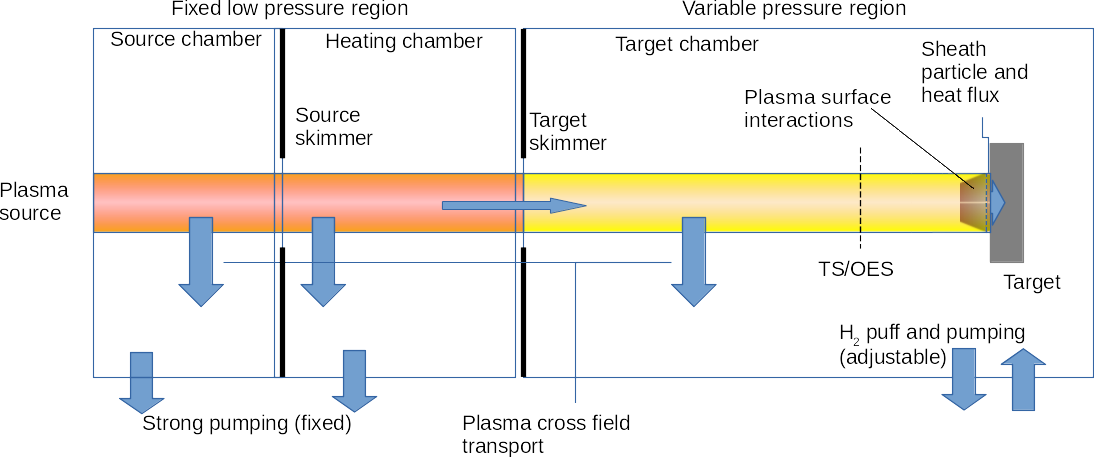}
	\caption{Schematic of the plasma column model used. This schematic is useful to correlate local properties (at TS/OES location) to global ones like the total input power. Simplifications as constant densities and temperatures along the magnetic field and constant flow speed are used.}
	\label{fig:plasma_column1}
\end{figure*}

To avoid considering precursor densities that could well match the line spectra but would lead to unrealistic power or particle losses a balance on the plasma column is performed. The definition of plasma column also allows to extract global information on the ELM-like pulse from the local TS/OES measurements and compare them with other global measurements like the power input from power supply. A schematic of the model of plasma column used is in \autoref{fig:plasma_column1}.

Fundamental assumptions are:
\begin{enumerate}
    \item Given that the neutrals density in source and heating chamber is low thanks to differential pumping, it is assumed that the plasma is transported undisturbed from the plasma source to the target skimmer. Here $T_e$,$n_e$ are equal to what is measured by TS in the target chamber for the lowest neutral pressure setting, that corresponds to the lowest possible volumetric losses.
    \item The plasma enters the target chamber without any molecular precursor. This is justified by the fact that from source to target chamber skimmer the neutral pressure is at it lowest while the temperature is at its highest and this conditions are the least favourable for reactions involving molecules.
    \item The neutral pressure is fixed throughout the ELM-like pulse at its steady state value.
    \item All plasma properties, such as temperatures, densities, reaction rates, radiated power, depend on radial and temporal coordinates only and are spatially constant from target skimmer to target. This is justified by the fact that the fast camera shows that in Stage 1 and 2 the radiation is mostly constant from a short distance off the target. Given the OES measuring location, only the properties of the bulk of the plasma can be analysed\footnote{Measurements specific to the region close to the target have been attempted but failed, possibly due to reflections or obstructions by the target itself.}. That means that the power losses in the visible light brightness peak between the OES and the target observed in Section III of \cite{Federici} cannot be accounted, so the volumetric power losses from the analysis will likely be an underestimation. The extent of the non uniform region close to the target, likely including the sheath and a region with strong plasma surface interactions, is typically <1cm, which is small compared to the 38 cm from target to skimmer, so this underestimation should be minor. Increasing neutral pressure from Stage 1 to 2 (the cases we are most interested in), the visible light brightness becomes stronger in the plasma column, making the anisotropy at the target even less relevant. The approximation also neglects\ anisotropy in the visible light brightness in the bulk for very high neutral pressure. This is especially dominant in Stage 3, so it's importance should be minor for Stages 1 and 2.
    \item The plasma behaviour in the sheath and in the region with strong plasma surface interactions is neglected.
    \item The flow velocity of the plasma is constant from the source to the target.
    \item Cross field transport is negligible (mostly true for charged particles due to the high magnetic field and additionally for molecular ions due to their short life time)
\end{enumerate}

Given these assumptions, one can calculate the components of the power and particle balance on the whole plasma column. The OES/TS measurements from a single location can be applied to the whole column and the contribution from atomic and molecular processes can be found.

A quantity that will be used later is the flow velocity ($v_{in}$), which is the velocity of the plasma as its flow from the target skimmer to the target. It is mainly used here to subdivide the power from the plasma source (a global value) to what is provided to each radial location and to estimate the local particle inflow. It is also used to estimate the kinetic energy of the plasma, but the relevance of this term is minor. There is no direct measurement of $v_{in}$ yet, as collective Compton scattering measurements will be available in the future. $v_{in}$ is then approximated by imposing, for the experimental condition with the lowest target chamber neutral pressure, that the power from the source matches the energy flow measured at the TS location. The flow is assumed having a single Mach number for all radial locations. Applying this conditions to \autoref{eq:plasma_column3} this translate to \autoref{eq:plasma_column1} that is then solved to find the Mach number.

\begin{equation}
\label{eq:plasma_column1}
\begin{aligned}
\sum_{r} \biggl\{ \biggl(\frac{1}{2} m_i v_{in}^2(r,t) +
&
5k_B T_{e,in}(r,t) + E_{ion} + E_{diss} \biggr) \cdot \phantom{\biggr\}} & \\ \phantom{\biggl\{} \cdot n_{e,in}(r,t) v_{in}(r,t)A \biggr\} &= P_{source}(t)
\\
v_{in}(r,t) &= M_{in}(t) c_{s,up}(r,t) \\  c_s &= \sqrt{\frac{ \left(T_e + T_{H^+{}}( {}\approx T_e) \right)k_B}{m_H} }
\end{aligned}
\end{equation}

In calculating $P_{source}$ as product of voltage and current an efficiency of 92\% is considered in the conversion from electric to plasma energy.\cite{Morgan2014} $M_{in}$ is $\approx$1 during the ELM-like pulse. It must be noted that at the beginning and end of the pulse TS is incapable of accurately measuring across the whole plasma because of the low density, and the energy conversion from electric power to plasma potential is lower, resulting in the calculated $M_{in}>1$. The effect of the overestimation, though, is to allow for larger energy and particle budgets, widening the possible parameter space, so it is acceptable.
I will now detail how to calculate the likelihood associated with the power and particle balance.

\subsection{Power balance}\label{Power balance}

In this chapter it will be detailed how the power (energy) balance equation is obtained and how all the terms are defined.
The 1D energy and particle balance equations are obtained from the 1D Fokker-Planck collisional kinetic equation (\autoref{eq:plasma_column2}) as per derivation from Stangeby\cite{Verhaegh2021} and are adapted using the mentioned approximations for the region from target skimmer to target. This results for every time step and radial location in \autoref{eq:plasma_column3}
\begin{equation}
\label{eq:plasma_column2}
\begin{aligned}
  \frac{ \partial f }{ \partial t}  + v_z \frac{\partial f}{\partial z}  + \frac{eE}{m} \frac{ \partial f }{ \partial v_z} = \left( \frac{ \partial f }{ \partial t} \right)_{coll}  +  S(x,v)
\end{aligned}
\end{equation}
\begin{equation}
\label{eq:plasma_column3}
\begin{aligned}
\frac{  \partial E }{ \partial t} +& \frac{d}{dz}  \left[ \left( \frac{1}{2} m_i v^2 + 5kT_e + E_{ion} + E_{diss} \right) n_e v \right] =\\ -& \underbrace{P_{ext\ source}}_{= 0} + P_{ volume\ sinks-sources }
\\
\frac{  \partial E }{ \partial t} +& \left( \frac{1}{2} m_i v_{in}^2 + 5kT_{e,in} + E_{ion} + E_{diss} \right) n_{e,in} v_{in} =\\ &  + P_{ target } + P_{ volume\ sinks-sources }
\\
P_{diss \: max} =& \underbrace{ \left( \frac{1}{2} m_i v_{in}^2 + 5kT_e + E_{ion} + E_{diss} \right)  {\frac{n_e V}{\Delta t}}}_{P_{ \partial t}} + \\ &+ \underbrace{ \left( \frac{1}{2} m_i v_{in}^2 + 5kT_{e,in} + E_{ion} + E_{diss} \right) n_{e,in} v_{in}A}_{P_{in}} \geq \\ \geq & P_{ volume\ sinks-sources }
\end{aligned}
\end{equation}
with $v$ the flow velocity, $E_{ion}$ and $E_{diss}$ the ionisation and dissociation energy for hydrogen. The inequality arises from not accounting the power delivered to the target and to neglect plasma interactions with neutrals such as elastic collisions and charge exchange. All quantities marked with the subscript ${}_{in}$ refer to the input conditions, otherwise the conditions inside the plasma column are intended. $P_{\partial t}$ represents the power deriving by depleting all the energy associated with the plasma in the volume of interest ($V$) in a single time step ($\Delta t$), $P_{in}$ is the power entering the volume of interest from the plasma source through the area $A$. $P_{diss \: max}$ is the maximum power that can be depleted in a radial portion of the plasma column in a time step. As an additional constrain on the power balance it will be required to $P_{volume\ sink-source}$ to be positive, as otherwise it would mean that the plasma is externally heated on its way to the target.

Let’s investigate now the volumetric sinks-sources term. There are roughly three ways in which a hydrogen plasma can undergo power losses:
\begin{enumerate}
    \item Radiative losses. This mostly comes from the relaxation of excited hydrogen atoms, which can arise from both plasma-atom as well as plasma-molecule interactions. The radiative losses associated with $H_2$ molecular band radiation are expected to be of insignificant.\cite{Groth2019} \label{Radiative losses}
    \item Power transfer from kinetic energy to potential energy. Several plasma species have a relative potential energy associated with it (for instance, $H^+$ has a potential energy of 13.6eV compared to atomic hydrogen). Converting a neutral into an ion thus “converts” 13.6eV of kinetic plasma energy to potential energy.  \label{Power transfer potential}
    \item Power transfer from CX and elastic collisions. CX as well as elastic collisions between the plasma and neutral atoms and molecules can lead to transfer of power from the plasma to the neutrals (and vice versa). This includes collisions between particles of the same specie but at different temperatures like the cold proton generated from ionisation and the hot one part of the plasma. \label{Power transfer CX}
\end{enumerate}

OES combined with collisional radiative models is used to estimate the magnitude of both path \ref{Radiative losses} and \ref{Power transfer potential}, which is employed in this work. For path \ref{Radiative losses}, the Balmer line emission is measured facilitating, through the Bayesian inference of the plasma properties, a full estimate of the hydrogenic line radiation from excited atoms arising both from plasma-atom as well as plasma-molecule interaction. For path \ref{Power transfer potential}, ionisation and recombination rates are estimated to account for the power transfer between potential and kinetic energy. In the recombination reaction a hot $H^+$ is converted into a neutral. That neutral has a kinetic energy equal to the temperature of the plasma that generated it, significantly higher than all other molecular and neutral species. For this reason the energy removed by the plasma assuming the neutrals from recombination escape is accounted in the local power balance. Additionally a series of molecular reactions are considered, see \autoref{Reactions} for which the difference in potential energy between reactants and products is calculated and accounted.

Note that paths \ref{Power transfer potential} and \ref{Power transfer CX} do not strictly represent power lost from the plasma column but can be power transfer mechanisms. Such transfer mechanisms often lead to an effective loss of kinetic energy by the plasma, but can also cause it to increase.

In the power balance that regards the limitation of the power transferred from the plasma at a single radial location, \autoref{eq:plasma_column3}, pathway \ref{Power transfer potential} is considered. It is in fact impossible for the plasma to transfer energy from kinetic to potential for more more than it is available. Differently when the quantity of interest are the components of the global power balance, \autoref{eq:plasma_column9}, only terms where the energy is removed from the plasma column entirely will be considered. Internal power transfer will not be considered as it is energy that remains in the plasma.

Pathway \ref{Power transfer CX} cannot be readily analysed experimentally but can be analysed in detail in simulations. The importance of this is currently discussed in literature and it could be significant especially in low temperature conditions in tokamak divertors and linear machines. \cite{Myatra2021,Smolders2020,Chandra2022} Further code investigations on this and detailed comparisons against experiments are required, which is outside of the scope of this work. To check that neglecting CX and $H_2$ elastic collisions, the ones to have the largest impact\cite{Chandra2022}, does not have a negative impact on the consistency of the solution a crude estimation was done in post processing. This is done by first calculating the ADAS CCD reaction rate for CX and AMJUEL 3.5 rate for $H_2$ elastic. These are multiplied by the density of the reactant species interested and by the maximum energy that can be transfer with a single collision. The energy of the reactants are equivalent to their temperature from TS and \ref{Priors from B2.5 Eunomia}. This results in \autoref{eq:CX_elastic}.
\begin{equation}
\label{eq:CX_elastic}
\begin{aligned}
  P_{CX} &= \frac{ 3 }{ 2}  (T_{H^+}(=T_e)-T_H) RR_{CX}(T_e,n_e) n_{H^+} n_{H}
  \\
  P_{H_2 elastic} &= \frac{ 3 }{ 2} \frac{8}{9} (T_{H^+}-T_{H_2}) RR_{H_2 elastic}(T_{H^+},T_{H_2})  n_{H^+} n_{H_2}
\end{aligned}
\end{equation}
These quantities PDFs are calculated as one of the outputs of the Bayesian algorithm.


The sinks/sources terms for \autoref{eq:plasma_column3} are taken from different sources, to encompass the best knowledge available at the time of writing, see \autoref{Reactions}. Grouping them by type and precursor the power balance sinks/sources term is then defined as per \autoref{eq:plasma_column4}

\begin{equation}
\label{eq:plasma_column4}
\begin{aligned}
P_{ \substack{volume \\ sinks-sources}} =& P_{ radiated } + P_{ \substack{neutral\ via \\ recombination} } + P_{ potential\ energy }
\\
P_{ radiated } =& P_{ radiated\ atomic } + P_{ radiated\ molecular } 
\\
P_{ \substack{neutral\ via \\ recombination} } =& \frac{3}{2} \Delta V T_e RR_{rec}
\\
P_{ potential\ energy } =& \Delta V \sum_{i} { {\Delta E}_i \cdot RR_i } 
\\
P_{ radiated\ atomic } =& \underbrace{P_{ excitation }}_{ADAS\ PLT} + \underbrace{P_{ rec + bremsstrahlung }}_{ADAS\ PRB}
\\
P_{ radiated\ molecular } =& P_{ rad\ {H_2}^+{} } + P_{ rad\ {H_2} } + P_{ rad\ {H}^-{}+{H_2}^+{} } + \\ &+ P_{ rad\ {H}^-{}+{H}^+{} } + P_{ rad\ e^-{} + H \rightarrow {H}^-{}+hv } 
\\
P_{ rad,i } =& \Delta V \sum_{p=2,q<p}^{p=13} \epsilon^{i}_{pq}
\end{aligned}
\end{equation}

where $\Delta V$ represent the volume corresponding to the radial location of the plasma considered, $\Delta E_i$ is the energy difference between products and reactants of the reaction $i$ and $RR_i$ is its reaction rate. The ${}^{p=13}$ comes from the fact that only atomic hydrogen excited states up to 13 are here considered. The probability that the inequality in \autoref{eq:plasma_column3} is true and that $P_{volume\ sinks-sources}$ is positive is calculated with \autoref{eq:plasma_column5}
\begin{equation}
\label{eq:plasma_column5}
\begin{aligned}
y =& P_{diss \: max} - P_{s-s}, \sigma_{y} = \sqrt{{\sigma_{P_{\substack{s-s}}}}^2 + {\sigma_{P_{diss \: max}}}^2 }
\\
L_P =& L(y \in [0,P_{diss \: max}]) = \\ =& \frac{1}{2} \left[ erf ( \frac{P_{diss \: max}-y}{ \sqrt{2} \sigma_{y} }) - erf(\frac{-y}{ \sqrt{2} \sigma_{y} } ) \right]
\end{aligned}
\end{equation}
where $P_{volume\ sinks-sources}$ is shortened with $P_{s-s}$.

The power sinks/sources are calculated by adding all the radiative losses to the potential energy contribution. The latter is itself composed by positive and negative contributions that tend to cancel out. This causes the uncertainty of the sinks/sources to greatly dominate over the input one, making the effective use of this balance very difficult. To solve this issue $\sigma_y$ considered as equal to $\sigma_{\substack{P_{diss \: max}}}$ and that is assumed to be 50\% of $P_{diss \: max}$ (using the fixed nominal $T_e$, $n_e$ values from TS).

\subsection{Particle balance}\label{Particle balance}

The derivation of the particle balance equation from Stangeby\cite{Stangeby2001} results in Equation 14

\begin{equation}
\label{eq:plasma_column6}
\begin{aligned}
\frac{ \partial n_j}{ \partial t} + \frac{d}{dz}   \left(n_j v \right) &= Sinks-Sources \\ (nv)_{j, diss \: max} &={\underbrace{\frac{n_j V }{ \Delta t }}_{(nv)_{j,\partial t}} + \underbrace{n_{j,in} v_{in}}_{(nv)_{j,in}}} \geq \\ \geq & (nv)_{j, Sinks-Sources} = \Delta V \sum_{i} f_{ij} RR_i \\ & \phantom{(nv)_{j, Sinks}}= \sum_{i} \dot{n}_{i,sinks} - \dot{n}_{i,sources}
\end{aligned}
\end{equation}

with $f_{ij}$ the multiplicity and sign in the reaction $i$ for the specie $j$ and ${n_{{H_2}}}_{in}={n_{{H}}}_{in}={n_{{H_2}^+}}_{in}={n_{H^-}}_{in} =0$. The inequality comes from not including the particles lost due to surface processes happening at the target and cross field transport. Charged particles are bound by magnetic fields while neutrals can more easily move across. For this reason the particle balance, that considers each radial location independently, is calculated only for charged particles as $e^-$, $H^+$, ${H_2}^+$, $H^-$.
In the case of ${H_2}^+$, $H^-$ the lifetime is very short so even in a single time step it is not physical to allow for its accumulation.\cite{Verhaegh2018} For this reason for them $(nv)_{i, Sinks-Sources}$ is limited to be $\geq -(nv)_{i, diss \: max}$. For $e^-$ and $H^+$ the net rate of production is limited to their density the next time step. This term, referred as $(nv)_{j,next}$, is defined similarly to $(nv)_{j,\partial t}$ in \autoref{eq:plasma_column6}. The likelihood that the particle balance is verified is given by \autoref{eq:plasma_column7a} and \ref{eq:plasma_column7b}

\begin{equation}
\label{eq:plasma_column7a}
\begin{aligned}
y_j &= (nv)_{j, diss \: max} - (nv)_{j, Sinks-Sources} \\ {\sigma}_{y_j} &=\sqrt{{{\sigma}_{(nv)_{j, diss \: max}}}^2 + {{\sigma}_{(nv)_{j, Sinks-Sources}}}^2 }
\\
L\left(y_{e^-{}} \in \right. & \left. [0,(nv)_{e^-, diss \: max}+(nv)_{e^-, next}]\right) =\\=& \frac{1}{2} \left[ erf \left(\frac{(nv)_{e^-, diss \: max}+(nv)_{e^-, next}-y_{e^-}}{\sqrt{2} {\sigma}_{y_{e^-}} } \right) \right. +\\ &-\left. erf \left( \frac{-y_{e^-{}}}{\sqrt{2} {\sigma}_{y_{e^-{}}} } \right) \right]
\\
L\left(y_{H^+{}} \in \right. & \left. [0,(nv)_{H^+{}, diss \: max}+(nv)_{H^+{}, next}]\right) =\\=& \frac{1}{2} \left[ erf \left(\frac{(nv)_{H^+{}, diss \: max}+(nv)_{H^+{}, next}-y_{H^+{}}}{\sqrt{2} {\sigma}_{y_{H^+{}}} } \right) \right. +\\ &-\left. erf \left( \frac{-y_{H^+{}}}{\sqrt{2} {\sigma}_{y_{H^+{}}} } \right) \right]
\end{aligned}
\end{equation}

\begin{equation}
\label{eq:plasma_column7b}
\begin{aligned}
L\left(y_{{H_2}^+{}} \in \right. & \left. [0,2(nv)_{{H_2}^+{}, diss \: max}]\right) = \\ =& \frac{1}{2} \left[ erf \left(\frac{2(nv)_{{H_2}^+{}, diss \: max}-y_{{H_2}^+{}}}{\sqrt{2} {\sigma}_{y_{{H_2}^+{}}} } \right) \right. +\\ &-\left. erf \left(\frac{-y_{{H_2}^+{}}}{\sqrt{2} {\sigma}_{y_{{H_2}^+{}}} } \right) \right]
\\
L\left(y_{{H}^-{}} \in \right. & \left. [0,2(nv)_{{H}^-{}, diss \: max}]\right) = \\ =& \frac{1}{2} \left[ erf\left(\frac{2(nv)_{{H}^-{}, diss \: max}-y_{{H}^-{}}}{\sqrt{2} {\sigma}_{y_{{H}^-{}}} } \right) \right. +\\ &- \left. erf\left(\frac{-y_{{H}^-{}}}{\sqrt{2} {\sigma}_{y_{{H}^-{}}} } \right) \right]
\\
L_{nv} &= L(y_{e^-{}} \geq 0) \cdot L(y_{H^+{}} \geq 0) \cdot \\ & \phantom{=} \cdot L(y_{{H_2}^+{}} \in [0,2(nv)_{{H_2}^+{}, diss \: max}]) \cdot \\ & \phantom{=} \cdot L(y_{{H}^-{}} \in [0,2(nv)_{{H}^-{}, diss \: max}]) 
\end{aligned}
\end{equation}

Similarly to what mentioned for the power balance, here too the sinks/sources term is composed of positive and negative factor, so rather than using it a large uncertainty on $(nv)_{i, diss \: max}$ of 100\% for $e$, $H^+$ (using the fixed nominal $n_e$ value from TS) and 5\% of TS $n_e$ for ${H_2}^+$, $H^-$ is assumed. The uncertainties are here adopted so large because differently from power and emissivity there is no direct measurement of the particle input.
As part of the particle balance one has also to include that the density of excited states obtained with ADAS and Yacora coefficients is lower than the density of total atomic hydrogen in the volume. This is calculated with \autoref{eq:plasma_column8}.

\begin{equation}
\label{eq:plasma_column8}
\begin{aligned}
y = n_{H} - \sum_{q,i} n_{i H(q)}, \sigma_{y} = \sqrt{ \sum_{q,i} (\sigma_i n_{i H(q)})^2  }
\\
L_{H^{exc}} = L(y \geq 0) = \frac{1}{2} \left[ 1 - erf\left(\frac{-{y}}{ \sqrt{2} \sigma_{y} } \right) \right]
\end{aligned}
\end{equation}

\subsection{Plasma column power balance}\label{Plasma column power balance}

The definition of the plasma column volume and the power balance allows to evaluate the global performance of the detached target to the ELM-like pulse. The parameter of interest is, in this case, how much power is removed from the plasma column and how much is due to atomic versus molecular effects.
In considering this the potential energy exchange due to EIR, for example, was not considered because it represent a transfer mechanism and not a net loss, while the radiative component due to radiation (ADAS PRB coefficient, returning the losses due to line radiation and Bremsstrahlung) was. Bremsstrahlung radiation is present also in the wavelength range of the IR camera and could be related to the observed prompt emission but this was not investigated. Similarly all other exchanges of potential energy are not considered. The generated neutrals while travelling out of the plasma can react with the neighbouring plasma and the energy they carry be reintroduced. Because evaluating this would require a significant effort this component is for now excluded.

The terms considered for the global power removed from the plasma column are indicated in \autoref{eq:plasma_column9}.
\begin{equation}
\label{eq:plasma_column9}
\begin{aligned}
 E_{ \substack{removed \\ from \\ plasma}} = E_{ radiated } =& \underbrace{E_{exc} + E_{rad\ rec+bremm}}_{atomic} +\\&+ E_{rad\ {H_2} } + E_{rad\ {H}} + E_{rad\ {H_2}^{+{}} } +\\& \underbrace{ +E_{rad\ {H}^{-{}} } + E_{rad\ e^-{}+H \rightarrow {H}^{-{}}+pv \phantom{+}}}_{ molecular }
\end{aligned}
\end{equation}
Once all the likelihoods associated with each combination of priors are calculated they are multiplied to return the total likelihood as per \autoref{eq:plasma_column10}

\begin{equation}
\label{eq:plasma_column10}
\begin{aligned}
L = P_{\epsilon} L_p L_{nv} L_{H^{exc}}
\end{aligned}
\end{equation}

The PDFs of all the components of the power and particle balance previously calculated are built by portioning their range to a smaller number of logarithmic intervals, summing the probability within. For each additive term of interest the PDFs are then convolved in space and in time to obtain the PDF for the whole ELM-like pulse. For each radial and time location and for each output required (for example the total radiated energy) is defined a large list of the possible energy losses in that section of the plasma. The list is randomly distributed according on the PDF of that output at that location. The contribution for all radii is summed to generate a list of possible values of the total radiated energy at one time step. A histogram is built based on that to represent the PDF of the quantity of interest at that time step. This operation is repeated to sum the contribution from all the time steps to return the PDF of the quantity of interest for the whole ELM-like pulse.

\subsection{Reactions}\label{Reactions}

\begin{table}
\begin{tabular}{ | p{5.5cm}| m{2.8cm} | } 
\hline
Reaction: & Chapter / type \\ 
\hline
$H^+ + e^- \rightarrow H(p) + h\nu$ \newline $H^+ + 2e^- \rightarrow H(p) + e^-$ & ADC,PRB,PEC \\ 
\hline
$H + e^- \rightarrow H^+ + 2e^-$ & SCD \\
\hline
$H(q) + e^- \rightarrow H(p>q) + e^-$ & PLT,PEC\\
\hline
\end{tabular}
\caption{Reactions whose rates and reference coefficients were sourced from the ADAS database.\cite{Summers2004,OMullane2013}}
\label{tab:adas}
\end{table}

\begin{table}
\begin{tabular}{ | p{5.5cm}| m{2.8cm} | } 
\hline
Reaction: & Chapter / type \\ 
\hline
${H_2}^+ + H^- \rightarrow H(p) + H_2$ &  \\
\hline
$H^+ + H^- \rightarrow H(p) + H$ & \\
\hline
${H_2}^+ + e^- \rightarrow H(p) + H(1)$ & \\
\hline
${H_2}^+ + e^- \rightarrow H(p) + H^+ + e^-$ & \\
\hline
$H_2 + e^- \rightarrow H(p) + H(1) + e$ & \\
\hline
\end{tabular}
\caption{Reactions whose rates and reference coefficients were sourced using the Yacora collisional radiative code.\cite{Wunderlich2016,Wunderlich2020}}
\label{tab:yacora}
\end{table}

\begin{table}
\begin{tabular}{ | p{5.5cm}| m{2.8cm} | } 
\hline
Reaction: & Chapter / type \\ 
\hline
$e^- + H_2 \rightarrow 2e^- + {H_2}^+$ & 4.11 Reaction 2.2.9 \\
\hline
$e^- + H_2 \rightarrow 2e^- + H + H^+$ & 4.12 Reaction 2.2.10 \\
\hline
$e^- + {H_2}^+ \rightarrow 2e^- + H^+ + H^+$ & 4.13 Reaction 2.2.11 \\
\hline
$e^- + {H_2}^+ \rightarrow e^- + H + H^+$ & 4.14 Reaction 2.2.12 \\
\hline
$e^- + {H_2}^+ \rightarrow H + H$ & 4.15 Reaction 2.2.14 \\
\hline
$e^- + H_2 \rightarrow e^- + H_2(v) \rightarrow H + H^-$ & 2.23 Reaction 2.2.17 \\
\hline
$H^+ + H_2(v) \rightarrow H + {H_2}^+$ & 3.28 Reaction 3.2.3 \\
\hline
$H^+ + H^- \rightarrow H + H$ & 4.52 Reaction 7.2.3a \\    
\hline
$e^- + H_2(v) \rightarrow e^- + H + H$ & 4.10 Reaction 2.2.5g \\    
\hline
\end{tabular}
\caption{Reactions whose rates and reference coefficients were sourced from the AMJUEL database.\cite{Reiter2017,Reiter2005,Kotov2007}}
\label{tab:amjuel}
\end{table}


\begin{table}
\begin{tabular}{ | p{5.5cm}| m{2.8cm} | } 
\hline
Reaction: & Chapter / type \\ 
\hline
${H_2}^+(vi) + H^- \rightarrow ({H_3}^*) \rightarrow $\newline $\rightarrow H_2(X1\Sigma_g;v_0) + H(n\geq 2)$ & 7.4.1 \\    
\hline
${H_2}^+(vi) + H^- \rightarrow ({H_3}^*) \rightarrow $\newline $\rightarrow H_2(N1,3\Lambda_{\sigma};v_0) + H(1),N \leq 4$ & 7.4.1 \\   
\hline
\end{tabular}
\caption{Reactions whose rates and reference coefficients were sourced from the Janev database.\cite{Janev2003}}
\label{tab:janev}
\end{table}

Reaction rates and other coefficients used in the Bayesian calculations are obtained from ADAS\cite{Summers2004,OMullane2013} for the atomic reactions while from Yacora \cite{Wunderlich2016,Wunderlich2020}, AMJUEL\cite{Reiter2017,Reiter2005,Kotov2007} 
or a collection of reaction rates from Janev\cite{Janev2003} for the molecular reactions. The reactions considered in this work are listed in \autoref{tab:adas}, \ref{tab:yacora}, \ref{tab:amjuel}, 
and \ref{tab:janev}.

\end{document}